\documentclass{aa}  
\usepackage{graphicx}
\usepackage{natbib}
\usepackage{float}
\usepackage{txfonts}
\begin{document}

   \title{Predicted MOND velocity dispersions for a catalog of ultra-diffuse galaxies in group environments}
\author{
          Oliver M\"uller\inst{1}
          \and
          Benoit Famaey\inst{1}
          \and 
          Hongsheng Zhao\inst{2}
          }

 \institute{Universit\'e de Strasbourg, Observatoire Astronomique de Strasbourg  (ObAS), CNRS UMR 7550 Strasbourg, France
\and
Scottish Universities’ Physics Alliance, University of St Andrews, St Andrews, UK. 
}

   \date{Received 18.12.2018; accepted 09.01.2019}

% \abstract{}{}{}{}{} 
% 5 {} token are mandatory
 
  \abstract
  % context heading (optional)
   %{} leave it empty if necessary  
   {The possibility that ultra-diffuse galaxies lacking dark matter has recently stimulated interest to check the validity of Modified Newton Dynamics (MOND) predictions on the scale of such galaxies.  It has been shown that the External Field Effect (EFE) induced by the close-by galaxy can suppress the velocity dispersion of these systems, so that they appear almost dark matter free in the Newtonian context. Here, following up on this, we are making {\it a priori} predictions for the velocity dispersion of 22  ultra-diffuse galaxies in the nearby Universe. This sample can be used to test MOND and the EFE with future follow-up measurements. We construct a catalog of nearby ultra-diffuse galaxies in galaxy group environments, and set upper and lower limits for the possible velocity dispersion allowed in MOND, taking into account possible variations in the mass-to-light ratio of the dwarf and in the distance to the galaxy group. The prediction for the velocity dispersion is made as a function of the three dimensional separation of the dwarf to its host. In 17 out of 22 cases, the EFE plays a crucial role in the prediction.}
  % aims heading (mandatory)
  % {}
  % methods heading (mandatory)
  % {}
  % results heading (mandatory)
  % {}
  % conclusions heading (optional), leave it empty if necessary 
  % {}


   \maketitle
%
%________________________________________________________________

\section{Introduction}
In recent years, the number of newly discovered ultra-diffuse galaxies (UDGs) in the nearby Universe has exploded. While it has long been known that this special type of galaxy exists in galaxy clusters environments \citep[e.g., ][]{1984AJ.....89..919S,1988ApJ...330..634I}, the discovery of UDGs in nearby galaxy groups \citep[e.g., ][]{2014ApJ...795L..35C,2016ApJ...833..168M,2018ApJ...868...96C} has brought new insights, as well as opportunities to better study these systems. Today, UDGs have been defined by their low surface brightness  of $>25$ mag arcsec$^{-2}$ in the $V$-band, and large radial extent, with projected half-light radii $r_{\rm eff}>1.5$\,kpc \citep{2015ApJ...798L..45V}. In morphology, they are not distinguishable from normal dwarf galaxies and appear both as early and late type galaxies \citep{1984AJ.....89..919S}.  Their formation is still under intense debate with different proposed formation mechanisms. For example it has been argued that these galaxies are failed Milky Way type galaxies \citep{2015ApJ...798L..45V,2015MNRAS.452..937Y}, just normal dwarf galaxies affected by stellar feedback, quenching, and outflows \citep{2016MNRAS.459L..51A,2017MNRAS.466L...1D,2018MNRAS.478..906C}{, or  rather a mixed bag of objects \citep{2018ApJ...856L..31T}.}

Recently, the UDG NGC\,1052-DF2, a putative member of the NGC\,1052 galaxy group \citep{2018ApJ...868...96C}, located at 20\,Mpc, has caught some attention, following the announcement by \citet{2018Natur.555..629V} that this galaxy is apparently lacking dark matter. This result was inferred from the velocity dispersion of ten bright globular clusters associated with this galaxy. Using a biweight estimator, \citet{2018Natur.555..629V} derived a velocity dispersion of only $\sigma_{vel}=3.2$\,km s$^{-1}$ with a 90\% upper limit of $\sigma_{vel} = 10.5$\,km s$^{-1}$, which is consistent with the baryonic matter of this galaxy alone. Following a more conservative approach, \citet{2018ApJ...859L...5M} obtained a value of  $\sigma_{vel}=9.2$\,km s$^{-1}$, with a 90\% upper limit of $\sigma_{vel} = 17.2$\,km s$^{-1}$. While the globular cluster population can be used as a tracer for the mass of the system {\citep{2018ApJ...856L..31T}}, ultimately, the velocity dispersion of the stellar body should be used. For example, in the Fornax dwarf spheroidal, where both estimates are available, there is a strong discrepancy between the two values \citep{2018MNRAS.tmp.2765L}. We note that, for NGC\,1052-DF2 a stellar velocity dispersion has recently been measured using IFU observations with VLT+MUSE { \citep{2018arXiv181207345E,2018arXiv181207346F}}.

The apparent lack of dark matter {in a dwarf galaxy} could have some challenging consequences for alternative gravity scenarios such as Modified Newton Dynamics (MOND, \citealt{1983ApJ...270..365M}), {as previously investigated by, e.g., \citet{2007A&A...472L..25G,2012A&A...543A..47G,2015A&A...584A.113L}.} The argument is that, in MOND, the baryonic matter alone gives the high observed velocities in galaxies, and notably the flat rotation curves in spirals. {This argument was indeed brought up by \citet{2018Natur.555..629V} in the case of NGC\,1052-DF2.} This simple assessment, however, is only true if the galaxy resides in relative isolation, that is, if no other galaxy is nearby. In the case that there is another gravitational potential close by, its environmental effect has to be taken into account when estimating the internal velocity dispersion, a phenomenon exclusively appearing in MOND. This so-called external field effect (EFE) arises from the non-linearity of the MONDian Poisson equation \citep{1984ApJ...286....7B}. 

The EFE can ultimately lower the velocity dispersion of a system, or in other words, make it Newtonian. Because this EFE appears only in MOND, and not within the standard model of cosmology, it is a highly intriguing feature to test and distinguish these paradigms. For a thorough discussion of the EFE and its consequences, see e.g. \citet{2016MNRAS.458.4172H,2018A&A...609A..44T}; the Supplementary Materials of \citet{kroupa2018does}; or the extensive review of MOND \citep{2012LRR....15...10F}.

For the Andromeda dwarf galaxies in the Local Group, it has been realized that this EFE plays a crucial role, allowing MOND to successfully predict the velocity dispersion of those dwarfs \citep{2013ApJ...766...22M,2013ApJ...775..139M}. For the ``feeble-giant" Crater-II \citep{2016MNRAS.459.2370T}, an ultra-faint dwarf galaxy satellite of the Milky Way,  a very low velocity dispersion has been predicted due to the EFE \citep{2016ApJ...832L...8M} and subsequently measured \citep{2017ApJ...839...20C}. Several more predictions of velocity dispersions for the ultra-faint dwarf galaxies around the Milky Way \citep{2014MNRAS.440..908P,2015MNRAS.453.1047P,2017MNRAS.470.1086C} now await follow-up measurements.  Taking the EFE into account for  NGC\,1052-DF2, the {expected} velocity dispersion was estimated to be  $\sigma_{vel} \simeq 14$\,km s$^{-1}$, consistently estimated by two independent teams \citep{2018MNRAS.480..473F,kroupa2018does}. While the former  used an analytic expression to get this results, the latter invoked an N-body integrator \citep{2009MSAIS..13...89L} to get this estimate, thereby validating the analytic estimate. This result alleviates the claim that NGC\,1052-DF2 could falsify MOND, but shows that UDGs are an excellent testbed to study the impact of the EFE. Following up on this previous work, we hereby make {\it a priori} predictions for the velocity dispersion of two dozen UDGs in the nearby Universe, thus {\it making MOND and its EFE falsifiable in a currently less-explored regime} than that of spiral galaxies. Interestingly, \citet{2018PhRvD..98j4036M,2018arXiv181112233M} has recently extended the succesful predictions of MOND to the regime of the internal dynamics of galaxy groups themselves.

The paper is organized as follows: In Section \ref{efe} we discuss how the velocity dispersion can be predicted in MOND, in Section \ref{data} we present a catalog of nearby UDGs in galaxy group environments,  in Section \ref{pred} we predict the velocity dispersion for the UDG catalog, and in Section \ref{discussion} we present a brief discussion and estimation on how these values can be measured. Finally, we give our summary and conclusions in Section \ref{summary}.

  \section{The External Field Effect calculation}
  \label{efe}
The Poisson equation of MOND reads, with \citet{2005MNRAS.363..603F} interpolating function: 
  
  \begin{equation}
     \nabla \cdot \left[{\mathbf g} \, \mu\left({g \over a_0}\right)\right] = 4 \pi G \rho, \quad \mu(x)={x \over 1+x},
\end{equation}
where ${\mathbf g}$ is the gravitational acceleration vector, $\rho$ the baryonic density, and $a_0=1.2\times10^{-13}$\,km s$^{-2}$ {\citep{1991MNRAS.249..523B}} is the MOND acceleration constant marking the transition from the Newtonian to the deep-MOND regime.
The fact that $\mu<1$ when in the regime of low acceleration ($g<a_0$), in which NGC\,1052-DF2 clearly is, would suggest that such galaxies should always display a `phantom' dark matter behaviour, i.e. a dark matter behavior when interpreted in the context of Newtonian dynamics.

However, when the galaxy is not isolated, one should take into account the EFE, such that
\begin{equation}
   {\mathbf g}={\mathbf g_{ex}} -\nabla \phi ,
\end{equation}
where ${\mathbf g_{ex}}$ is the gravitational field from the neighbouring galaxies, and $\phi$ the internal MOND potential.

 In a MONDian universe, the velocity dispersion of an {\it isolated} spherical and isotropic system in the low acceleration regime is easily calculated from its baryonic mass $M$ alone \citep[e.g.,][]{2013ApJ...766...22M}:
 \begin{equation}
      \sigma_{iso}=\Bigg(\frac{4}{81}G M a_0\Bigg)^{1/4},
 \end{equation}
 where M is the total baryonic mass.
 %$G=1.3276\times10^{11}$\,M$_\odot^{-1}$ km s$^{-2}$ is the gravitational constant,
 This formula applies when the total acceleration  (internal plus external acceleration) is well bellow $a_0$, i.e. $g+g_{ex}<a_0$. 
 
 When the internal acceleration is larger than the external,  $g_{ex} < g < a_0$, then we can apply the given formula for the isolated case. However, if  $ g < g_{ex} < a_0$ the object is quasi-Newtonian, where Newton's law of Gravity applies with a renormalization of the gravitational constant $G$ \citep{2013ApJ...766...22M,2018MNRAS.480..473F}. 
 The external acceleration can be approximated by $g_e=v_{rot}^2/D$, where $v_{rot}$ is the rotation curve of the external galaxy and $D$ is the separation between the two objects. Using Eq.\,59 of \citet{2012LRR....15...10F}, one can estimate the MOND acceleration $g$ at the half-light radius with:
  \begin{align}
 (g+g_{ex})\,\mu \Bigg(\frac{g+g_{ex}}{a_o}\Bigg) =g_{N,1/2} +g_{ex}\,\mu\Bigg(\frac{g_{ex}}{a_0}\Bigg), 
 \end{align}
 with
 \begin{align}
 g_{N,1/2}={G M \over 2 r_{1/2}^2},
  \end{align}
 where $\mu(x)=x/(1+x)$ is the simple interpolation function \citep{2005MNRAS.363..603F}, and $g_{N,1/2}$ is the Newtonian internal field for the mass embedded within the 3D deprojected half light radius ($r_{1/2}=4/3\,r_{\rm eff}$). The expression solved for $g$ is rather long, so we refrain from showing it here and refer to the footnote 34 of \citet{2012LRR....15...10F}, where it is explicitly given. We can now estimate the true velocity dispersion of the system, corrected for the external field. Having estimated $g$, we can calculate the renormalization of the gravitational constant with 
    \begin{align}
G_{renorm}=G\cdot(g/g_{N,1/2}). % I have change it from G_eff to G_renorm because eff could confuse with it being at the effective radius, which is not correct. 
\end{align}
  Finally, we can use the mass estimator in \citet{2010MNRAS.406.1220W} Eq. 2 to calculate the velocity dispersion:
  %$$\sigma=\sqrt{\frac{0.5M\,G_{\rm eff}}{4\,r_{\rm eff}}},$$
    \begin{align}
            \sigma=\sqrt{\frac{0.5M\,G_{renorm}}{3\,r_{1/2}}},
    \end{align}
where $0.5M$ corresponds to the mass embedded within $r_{1/2}$. This cooking recipe to calculate the velocity dispersion in the quasi-Newtonian regime is discussed in more detail in \citet{2018MNRAS.480..473F}, who used it to successfully derive the MONDian value for NGC\,1052-DF2 to be between 8.9 and 19.0 km s$^{-1}$ depending on the interpolating function, stellar mass-to-light ratio, and three-dimensional distance to the host. Note that the latest measured velocity dispersion value, using deep MUSE IFU observations \citep{2018arXiv181207345E} for the stellar body of NGC\,1052-DF2 is $\sigma_{stellar}=18.8$\,km s$^{-1}$, and for the globular cluster and planetary nebulae population associated to this galaxy is $\sigma_{GC,PN}=10.5$\,km s$^{-1}$ -- which are well within the allowed MONDian range for this system.

\section{A catalog of nearby ultra-diffuse galaxies}
\label{data}

\subsection{Data}

 Several independent teams have taken up the effort to search for nearby low-surface brightness galaxies in group  \citep[e.g., ][]{2014ApJ...787L..37M,2016ApJ...828L...5C,2016DGSAT,2017ApJ...850..109B,2015A&A...583A..79M,2017A&A...597A...7M,2018ApJ...863L...7M} {and cluster \citep[e.g.][]{2015ApJ...809L..21M,2016A&A...590A..20V,2017ApJ...834...16M,2017A&A...608A.142V,2018ApJ...855..142E} environments}. While most detections in these surveys correspond to the regime of the normal dwarf galaxies, also several UDGs have been discovered, based on their integrated light profiles. Up to now, no compilation of these objects exists. Therefore we have collected UDGs from the literature -- mainly selecting objects with half-light radius $r_{\rm eff}$ estimates larger than 1.5\,kpc (including those objects with very close values to 1.5\,kpc, e.g. NGC\,7814-DGSAT-7 with $r_{\rm eff}=1.49$\,kpc) and residing in a galaxy group environment based on their projected position in the sky. Therefore we exclude UDGs residing in clusters, because in these environments too many things can affect the velocity dispersion of the UDGs. Also, we only consider galaxies in the nearby Universe ($z<0.01$). While going further would indeed increase the sample \citep{2018ApJ...857..104G}, the galaxy's association to a certain galaxy group becomes more difficult to estimate.

 \begin{table*}[ht]
\caption{A list of nearby ultra-diffuse galaxies within galaxy groups.}% title of Table
     % is used to refer this table in the text
\setlength{\tabcolsep}{4pt} %
\centering                          % used for centering table
\begin{tabular}{l l l l l l l l l l}        % centered columns (4 columns)
\hline\hline                 % inserts double horizontal lines$\alpha_{2000}$ & $\delta_{2000}$ & Observing & Instrument & Exposure & Filter & Airmass & Image quality  \\    % table heading 
Name & RA & dec &	$m_V$ &	$\mu_{\rm eff,V}$ &	$r_{\rm eff}$  &	$\Delta_{proj}$ &	$(m-M)$ &  $L_\odot$&	ref \\    % table heading 
 & J2000 & J2000 & mag & mag arcsec$^{-2}$ & kpc & kpc & mag & 10$^6$  & \\ 
\hline      \\[-2mm]                  % inserts single horizontal line
NGC\,7814-DGSAT-2 & 00:03:06.9 & $+$16:18:30.8 & 17.9 & 26.9 & 2.29 & 47 & 31.02 & 30.1 & \citet{2017DGSAT} \\
NGC\,7814-DGSAT-7 & 00:00:44.0 & $+$15:27:14.3 & 18.7 & 27.0 & 1.49 & 258 & 31.02 & 14.4 & \citet{2017DGSAT} \\
NGC\,1052-DF4 & 02:39:15.1 & $-$08:06:58.6 & 16.5 & 25.1 & 1.60 & 165 & 31.51 & 171.8 & \citet{2018ApJ...868...96C} \\
NGC\,1052-DF1 & 02:40:04.6 & $-$08:26:44.4 & 18.2 & 27.4 & 2.51 & 109 & 31.51 & 35.9 & \citet{2018ApJ...868...96C} \\
NGC\,1052-DF2 & 02:41:46.8 & $-$08:24:09.3 & 16.2 & 25.1 & 2.06 & 79 & 31.51 & 226.5 & \citet{2018ApJ...868...96C} \\
NGC\,1084-DF1 & 02:42:38.0 & $-$07:20:16.3 & 16.2 & 24.7 & 1.63 & 280 & 31.34 & 193.8 & \citet{2018ApJ...868...96C} \\
NGC\,2683-DGSAT-1 & 08:52:47.8 & $+$33:47:33.1 & 14.7 & 26.5 & 4.10 & 68 & 30.05 & 234.5 & \citet{2016DGSAT} \\
NGC\,2683-DGSAT-2 & 08:55:23.3 & $+$33:33:32.4 & 16.3 & 25.8 & 1.39 & 105 & 30.05 & 53.7 & \citet{2016DGSAT} \\
M96-DF6 & 10:46:53.1 & $+$12:44:33.5 & 16.6 & 27.0 & 2.11 & 172 & 30.15 & 44.9 & \citet{2018ApJ...868...96C} \\
dw1055+11 & 10:55:43.5 & $+$11:58:05.0 & 16.9 & 26.7 & 1.78 & 411 & 30.15 & 34.0 & \citet{2018LeoI} \\
dw1117+15 & 11:17:02.1 & $+$15:10:17.0 & 17.4 & 27.4 & 2.04 & 398 & 30.15 & 21.5 & \citet{2018LeoI}\\
NGC\,3625-DGSAT-2 & 11:21:22.9 & $+$57:34:50.1 & 20.3 & 26.9 & 1.54 & 151 & 32.86 & 18.0 & \citet{2017DGSAT} \\
NGC\,3625-DGSAT-4 & 11:21:40.8 & $+$57 24 37.0 & 18.5 & 26.3 & 2.17 & 263 & 32.86 & 94.6 & \citet{2017DGSAT} \\
NGC\,3625-DGSAT-3 & 11:22:12.0 & $+$58:02:11.9 & 19.0 & 25.5 & 1.48 & 221 & 32.86 & 59.7 & \citet{2017DGSAT} \\
NGC\,3669-DGSAT-2 & 11:24:48.3 & $+$57:37:58.0 & 20.0 & 26.2 & 1.86 & 93 & 33.17 & 31.4 & \citet{2017DGSAT} \\
NGC\,3669-DGSAT-3 & 11:26:38.8 & $+$57:41:19.1 & 18.3 & 26.9 & 3.91 & 123 & 33.17 & 150.4 & \citet{2017DGSAT} \\
NGC\,4594-DGSAT-1 & 12:39:55.1 & $-$11:44:38.4 & 16.2 & 26.1 & 1.67 & 24 & 30.12 & 62.8 & \citet{2016DGSAT} \\
CenA-MM-Dw1 & 13:30:14.3 & $-$41:53:34.8 & 14.2 & 25.4 & 1.82 & 92 & 27.84 & 48.9 & \citet{2018arXiv180905103C} \\
M101-DF5 & 14:04:28.1 & $+$55:37:00.0 & 17.7 & 27.7 & 4.90 & 342 & 32.16 & 103.0 & \citet{2016ApJ...833..168M} \\
M101-DF7 & 	14:05:47.5 & $+$55:07:57.3 & 20.1 & 28.4 & 2.60 & 112 & 32.16 & 11.3 & \citet{2016ApJ...833..168M} \\
M101-DF4 & 	14:07:33.8 & $+$54:42:39.2 & 18.5 & 27.6 & 3.60 & 140 & 32.16 & 49.3 & \citet{2016ApJ...833..168M} \\
M101-DF6 & 14:08:18.7 & $+$55:11:30.6 & 19.8 & 27.5 & 2.90 & 117 & 32.16 & 14.9 & \citet{2016ApJ...833..168M} \\
\hline
\end{tabular}
\label{dataTable} 
\end{table*}

 In Table \ref{dataTable}, we present our sample of nearby UDGs within galaxy group environments. We have compiled the relevant data needed to estimate the velocity dispersion in MOND (see Section \ref{efe}). For this, we need the total luminosity of the galaxy $L_\odot$, the half-light radius $r_{\rm eff}$, and the projected separation to the host galaxy $\Delta_{proj}$.
 When missing, we have derived values from the published data, e.g. the mean surface brightness from the apparent magnitude and the half-light radius, the luminosity in solar units from the absolute magnitude, and the projected separation simply from the coordinates. We have transformed the apparent magnitudes into solar luminosities by using
 $L_\odot=10^{0.4\,(m_\odot-m+DM)}$,
 where $m_\odot=4.83$\,mag is the solar apparent magnitude in the $V$ band, $m$  is the apparent magnitude of the UDG, and $DM$ is the distance modulus to the host. 
 
 We also need the rotation curve of the host galaxies to estimate their gravitational influence. In our sample, galaxies with an asymptotic circular velocity measurement are: NGC\,1052 ($v_{rot}=210$\,km s$^{-1}$, \citealt{1986AJ.....91..791V}), M96 ($v_{rot}=$240\,km s$^{-1}$, \citealt{2004A&A...421..433M}), NGC\,3628 ($v_{rot}=$213\,km s$^{-1}$, \citealt{1993MNRAS.263.1075W})
 NGC\,1084 ($v_{rot}=140$\,km s$^{-1}$, \citealt{1963ApJ...137..376B}), NGC\,2683 ($v_{rot}=215$\,km s$^{-1}$, \citealt{1991AJ....101.1231C}), NGC\,4594/M\,104 ($v_{rot}=350$\,km s$^{-1}$, \citealt{1977ApJ...214..383F}), NGC\,7814 ($v_{rot}=215$\,km s$^{-1}$, \citealt{2011A&A...531A..64F}). {From there, the external field at the position of the UDG can be estimated directly from the centripetal acceleration.}
 
 For elliptical galaxies where no rotation curve data is available, we converted the $K$ band luminosity using a Mass-to-Light ($M/L$) ratio of 0.8 \citep{2008AJ....136.2648D} to get the total stellar mass. {From there, the gravitational field at the distance of the UDG can be estimated in the context of MOND.} This was the case for  Cen\,A ($K=-23.9$\,mag),  NGC\,3625 ($K=-21.9$\,mag), NGC\,3669 ($K=-22.3$\,mag),  NGC\,7814 ($K=-24.0$\,mag), NGC 5485 ($K=-23.8$\,mag), NGC\,3384 ($K=-23.4$\,mag), M\,105 ($K=-23.9$\,mag), NGC\,5475 ($K=-22.8$\,mag), and NGC\,3619 ($K=-24.0$\,mag) using the photometry by the Two Micron All Sky Survey \citep[2MASS, ][]{2006AJ....131.1163S} and a distance modulus given in Table \ref{dataTable}. 
 
 Finally, for one face-on spiral galaxy (NGC\,1042), no rotation curve was available. Therefore we first converted again its $K$ band luminosity using a Mass-to-Light ($M/L$) ratio of 0.8 to get the stellar mass of the system. Then, we estimated the additional total gas mass with Eq. 2 from \citet{2016MNRAS.456L.127D} and added this to the stellar mass {to get the total baryonic mass, and estimate the MOND external field at the position of the UDG from there}.
 
 Because galaxy group environments can host several large galaxies ($M_V<-19$\,mag) we have to be careful to select the influencing galaxy when calculating the MONDian velocity dispersion of the UDG. If there is a bright galaxy apart from the dominant galaxy in the group -- which is mostly indicated by the group's name -- close to the UDG we have to re-evaluate the MONDian velocity dispersion for those systems. A precise evaluation of the EFE as a function of the 3D position of the UDG in the group should then be performed using a numerical Poisson solver. In the present paper, we present an order of magnitude of the effect on the velocity dispersion by considering the EFE associated to each bright galaxy separately.
 
 \subsection{Individual galaxy groups}
 
 We now have a closer look at the different galaxy groups and evaluate the impact of their different bright galaxies onto the UDGs.
  
  {NGC\,1052 group:} This galaxy group consists of three bright galaxies, the dominant elliptical galaxy NGC\,1052, the face-on spiral galaxy NGC\,1042, and the edge-on spiral galaxy NGC\,1035.
  While NGC\,1052-DF2 is only affected by the dominant elliptical, we note that NGC\,1052-DF1 is at a separation of only 28\,kpc from NGC\,1042, which indeed has a major impact on the predicted velocity dispersion. On the other hand, NGC\,1052-DF4 is separated by only 21\,kpc from NGC\,1035.
  %($K=-22.4$\,mag at 20\,Mpc).
  
    {NGC\,1084 group:} To the north-east of the NGC\,1052 group, a single spiral galaxy -- NGC\,1084 -- resides well in isolation, bringing no additional complications to our calculations for NGC\,1084-DF1.
  
  {M\,96/Leo-I group:} This group consists of 7 bright galaxies (see e.g. Fig.\,1 in \citealt{2018LeoI}) at a mean distance of 10.7\,Mpc. Of those, the two elliptical galaxies M\,105 and NGC\,3384 have separations of 53\,kpc and 67\,kpc from M\,96-DF6, respectively. The large spiral galaxy M\,96 has a projected separation of 172\,kpc. This UDG sits almost at the center of this massive group, complicating accurate predictions. Fortunately, for M\,96-DF6, HST distance measurements are available \citep[$D=10.2\pm0.3$\,Mpc][]{2018ApJ...868...96C}, indicating that this UDG is at the nominal distance of M\,96 \citep[$D=10.4\pm0.3$\,Mpc][]{2013AJ....145..101K}, and farther separated from M\,105 \citep[$D=11.3\pm0.1$\,Mpc][]{2013AJ....145..101K} and NGC\,3384 \citep[$D=9.4\pm0.1$\,Mpc][]{2013AJ....145..101K}. 
  Farther outside of this galaxy assembly resides dw1055+11. 
  
  {Leo-Triplet:} The Leo-Triplet is sometimes considered as part of the Leo-I group, based on their same velocity and distance measurements. However, they are well separated on-sky into two distinct associations \citep{2018LeoI}. The dominant edge-on spiral galaxy NGC\,3628 is also the closest galaxy to the UDG dw1117+15, which is almost 2 degrees to the north of the system.

 {Cen\,A group:} Cen\,A is the closest elliptical galaxy in the nearby universe and is the dominant galaxy in the Cen\,A group \citep{2017A&A...597A...7M}. Because the distances to Cen\,A and Cen\,A-MM-dw1 are well established with high-precision tip of the red giant branch measurements \citep{2004A&A...413..903R,2018arXiv180905103C}, no confusion is expected for their association.
 
 {NGC\,5485 group:} Several surveys have targeted the nearby M\,101 group of galaxies and announced the discovery of new dwarf galaxy members \citep{2014ApJ...787L..37M,2017A&A...602A.119M,2017ApJ...850..109B}. Within its projected virial radius, reside the two massive elliptical galaxies -- NGC\,5485 and NGC\,5473 with similar mass estimates -- at a distance of 27\,Mpc. Some of the putative M\,101 members have recently been identified as background UDG members of the NGC\,5485 group by their non-detected red giant branch star population in HST observations \citep{2016ApJ...833..168M}. Many more candidates await distance measurements and can potentially be UDG members of the NGC\,5485 group as well. While M\,101-DF4, M\,101-DF6, and M\,101-DF7 are in projection closer to NGC\,5485 such that it is reasonable to use this galaxy as the major external field provider, the case of M\,101-DF5 is less clear. It has an on-sky separation of 44\,arcmin to NGC\,5485, and only 10\,arcmin and 23\,arcmin to the other nearby galaxies NGC\,5475 and 
NGC\,5443, respectively. These two galaxies have the same systemic velocity measurement as the NGC\,5485 group, which could indicate that M\,101-DF5 could be associated to these these two galaxies, and not to NGC\,5485 itself. We therefore consider two cases for M\,101-DF5, a) one case where it resides within NGC\,5485 external field, and b) another where it resides within NGC\,5475 external field.
 
 {NGC\,2683 group:} This galaxy group is made up of one large spiral galaxy and resides in isolation, making the predictions for NGC\,2683-DGSAT-1 and NGC\,2683-DGSAT-2 straightforward.
 
 {NGC\,4594/M\,104 group:} This famous galaxy group with the M\,104 (the Sombrero galaxy) in its heart is again, very isolated, making the prediction for NGC\,4594-DGSAT-1 straightforward.
 
 {NGC\,3625 group:} The isolated spiral galaxy NGC\,3625 is in projection very close to the shell galaxy NGC\,3619, and the elliptical galaxy NGC\,3613 which are well separated in velocity space from NGC\,3625. It is therefore difficult to disentangle the membership of the UDGs between those groups. If NGC\,3625-DGSAT-2 is a NGC\,3625 member, the external field of NGC\,3625 is almost negligible, however, if it is associated to NGC\,3619 ($D=25\pm1.8$\,Mpc), it is deep in the quasi-Newtonian regime. For the other two UDGs in this region (NGC\,3625-DGSAT-3, NGC\,3625-DGSAT-4), the separation to NGC\,3619 is too large to be strongly affect by the EFE.
 
  {NGC\,3669 group:} The isolated spiral galaxy NGC\,3669 is to the east of the NGC\,3625 and NGC\,3619 groups. However, their separation is too large for major confusions, allowing straightforward predictions for NGC\,3669-DGSAT-2 and NGC\,3669-DGSAT-3. 
  
  {NGC\,7184 group:} This group again is well isolated, with the spiral galaxy NGC\,7184 at its center, making the predictions for NGC\,7814-DGSAT-2 and .NGC\,7814-DGSAT-7 straightforward.
 
  \section{Predicted velocity dispersions}
  \label{pred}
  For all galaxies of Table \ref{dataTable} we have first calculated whether the external field dominates the dynamics or not. To make a distinction between the isolated and the quasi-Newtonian regimes we estimate the fraction $g/g_{ex}$ as a function of the three dimensional separation of the UDG to its main host (or to various possible hosts), up to one Megaparsec. We can safely assume that, when the internal acceleration is less than twice the external one, the EFE will play a role. In this case, we thus use the $\sigma$ as previously derived by including the EFE in Eqs. 4-7. Above this value of the internal to external field ratio, the galaxy can be assumed to be isolated, therefore $\sigma_{iso}$ applies as in Eq. 3.

\begin{figure*}[ht]
\includegraphics[width=4.5cm]{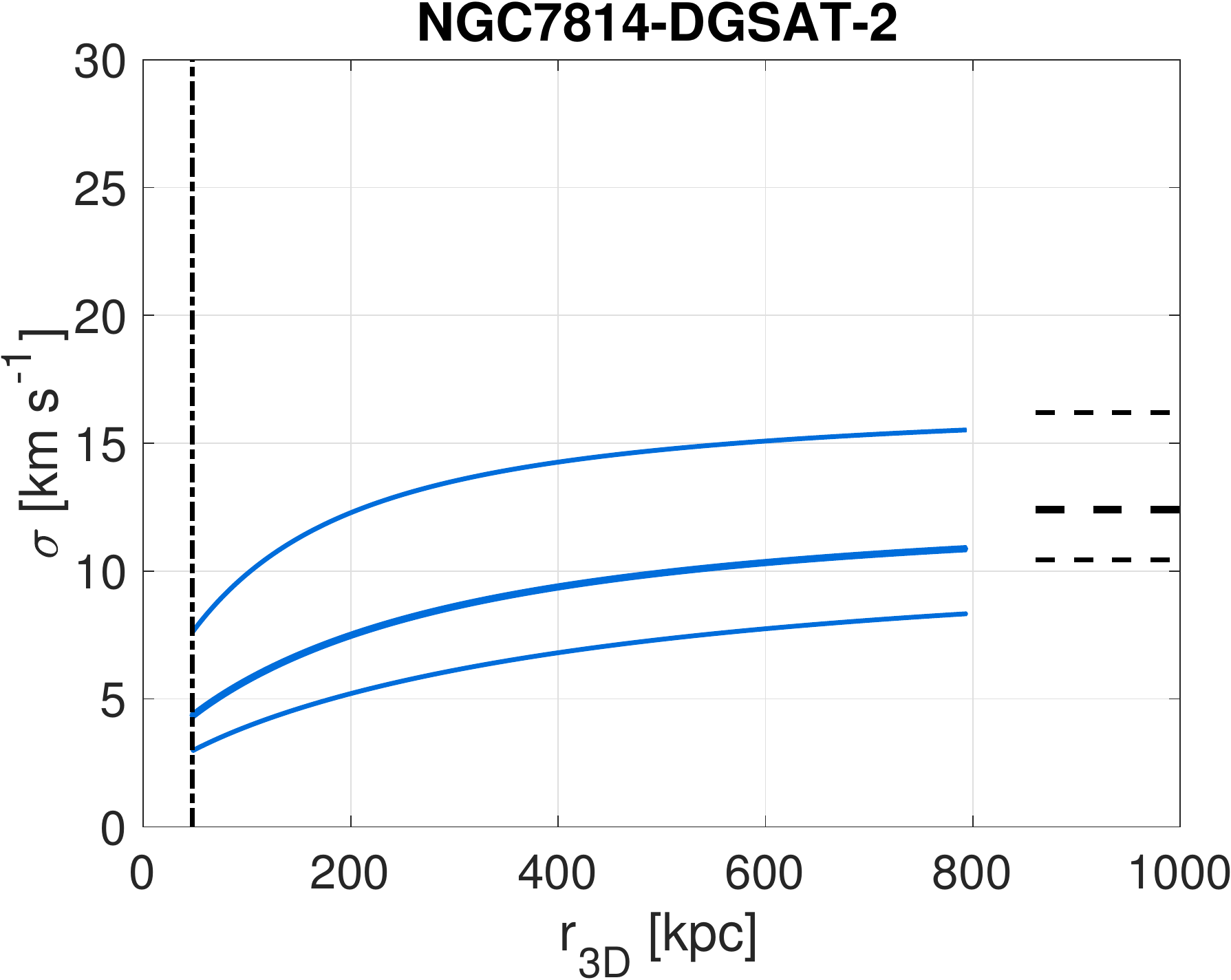}
\includegraphics[width=4.5cm]{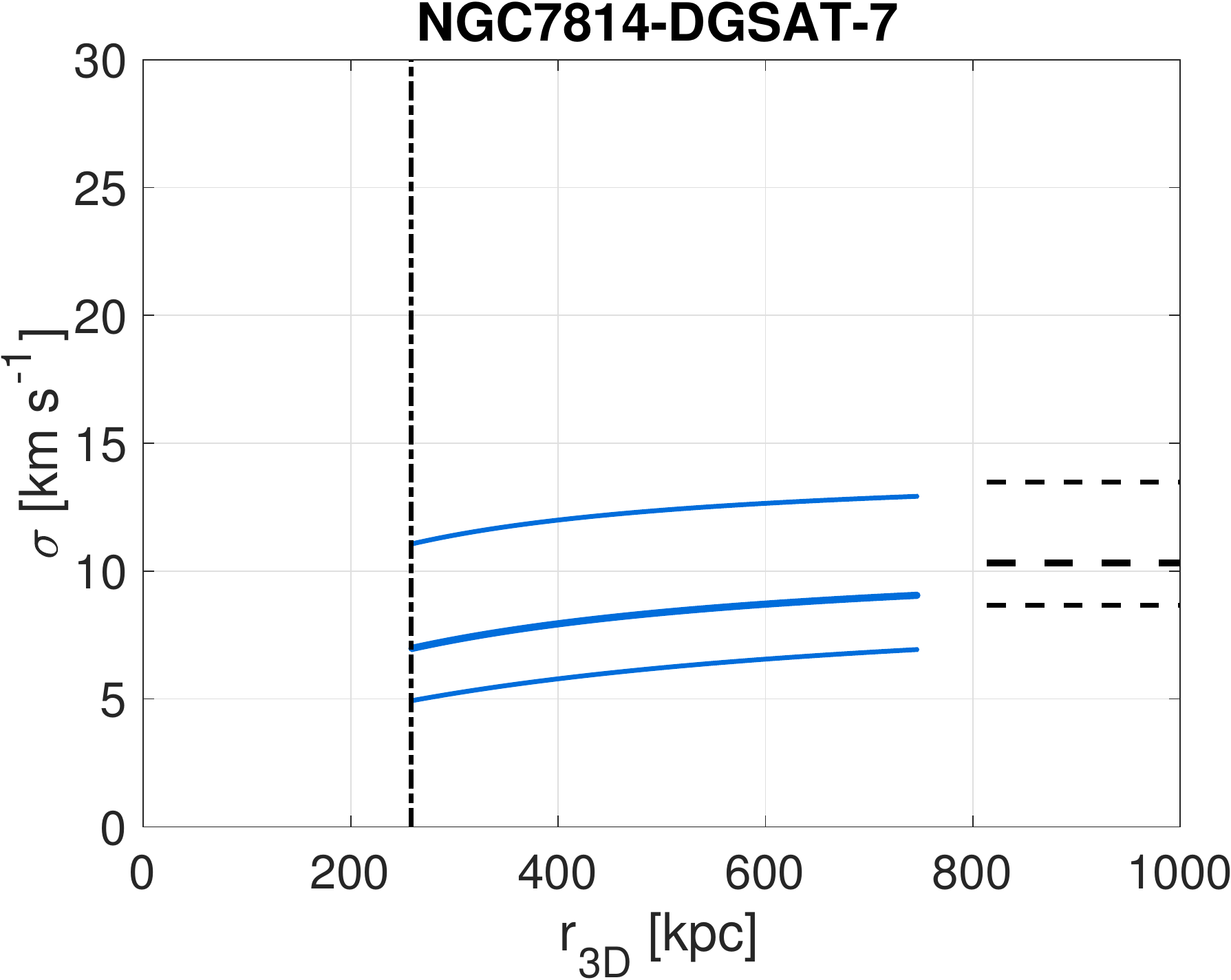}
\includegraphics[width=4.5cm]{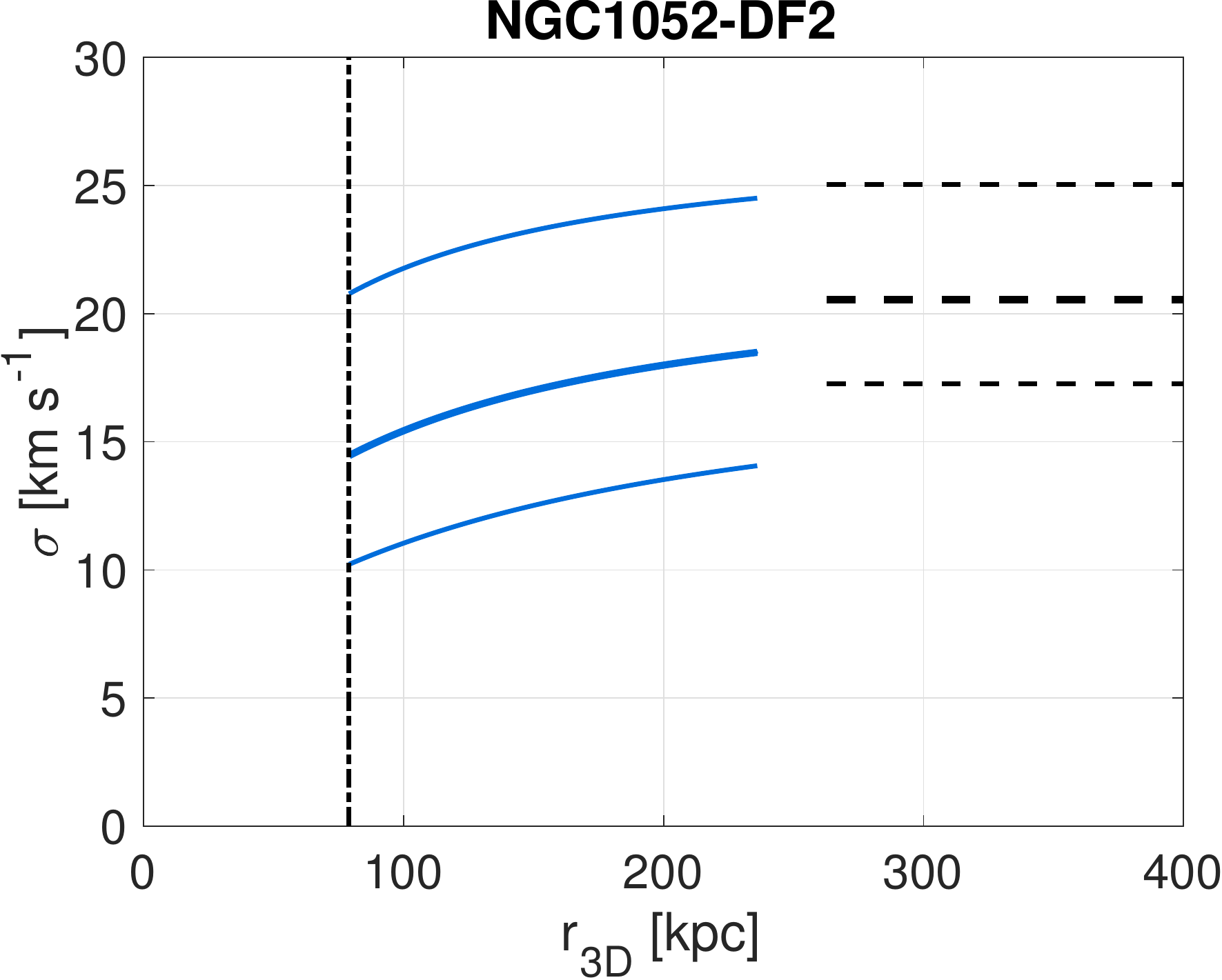}
\includegraphics[width=4.5cm]{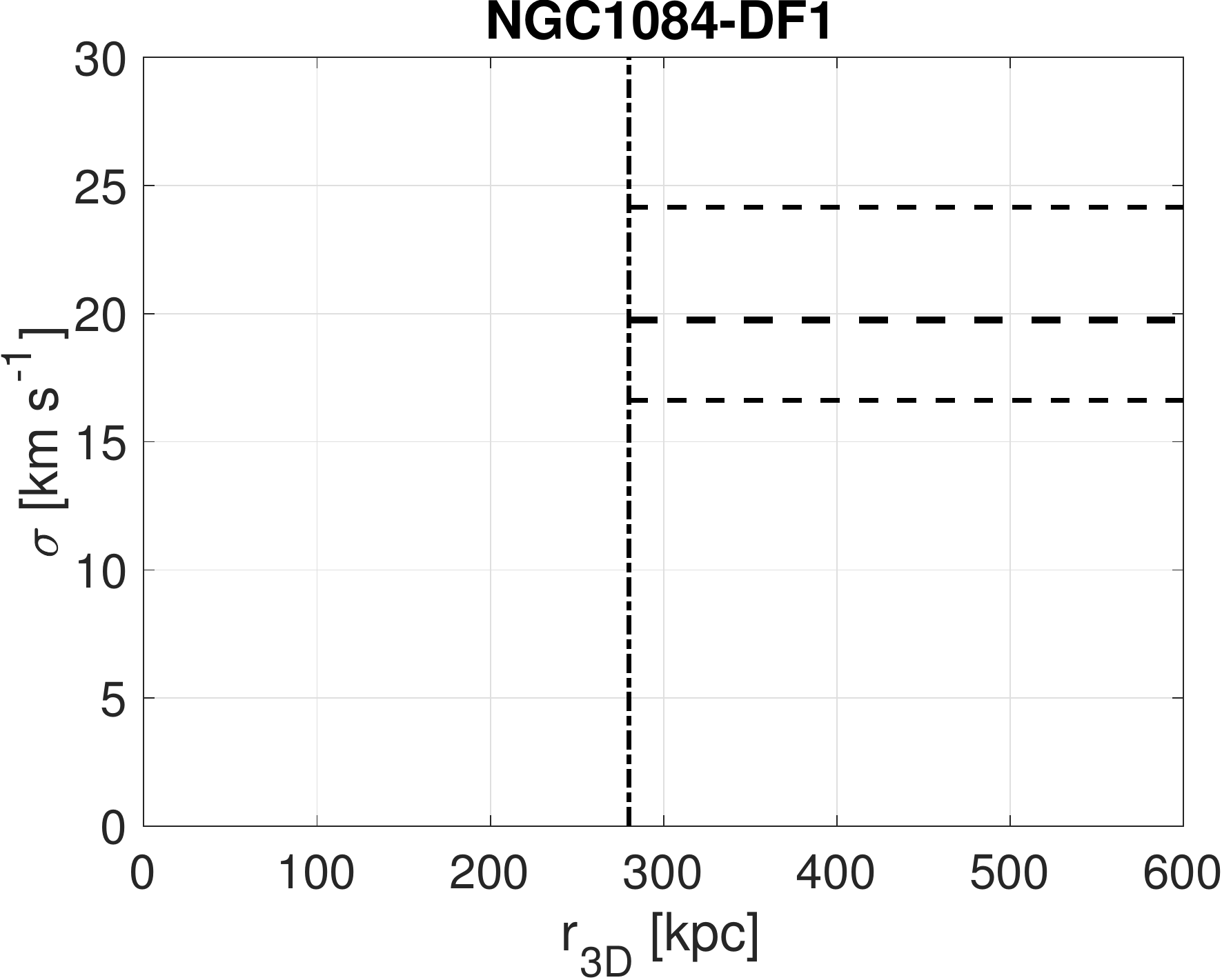}\\
\includegraphics[width=4.5cm]{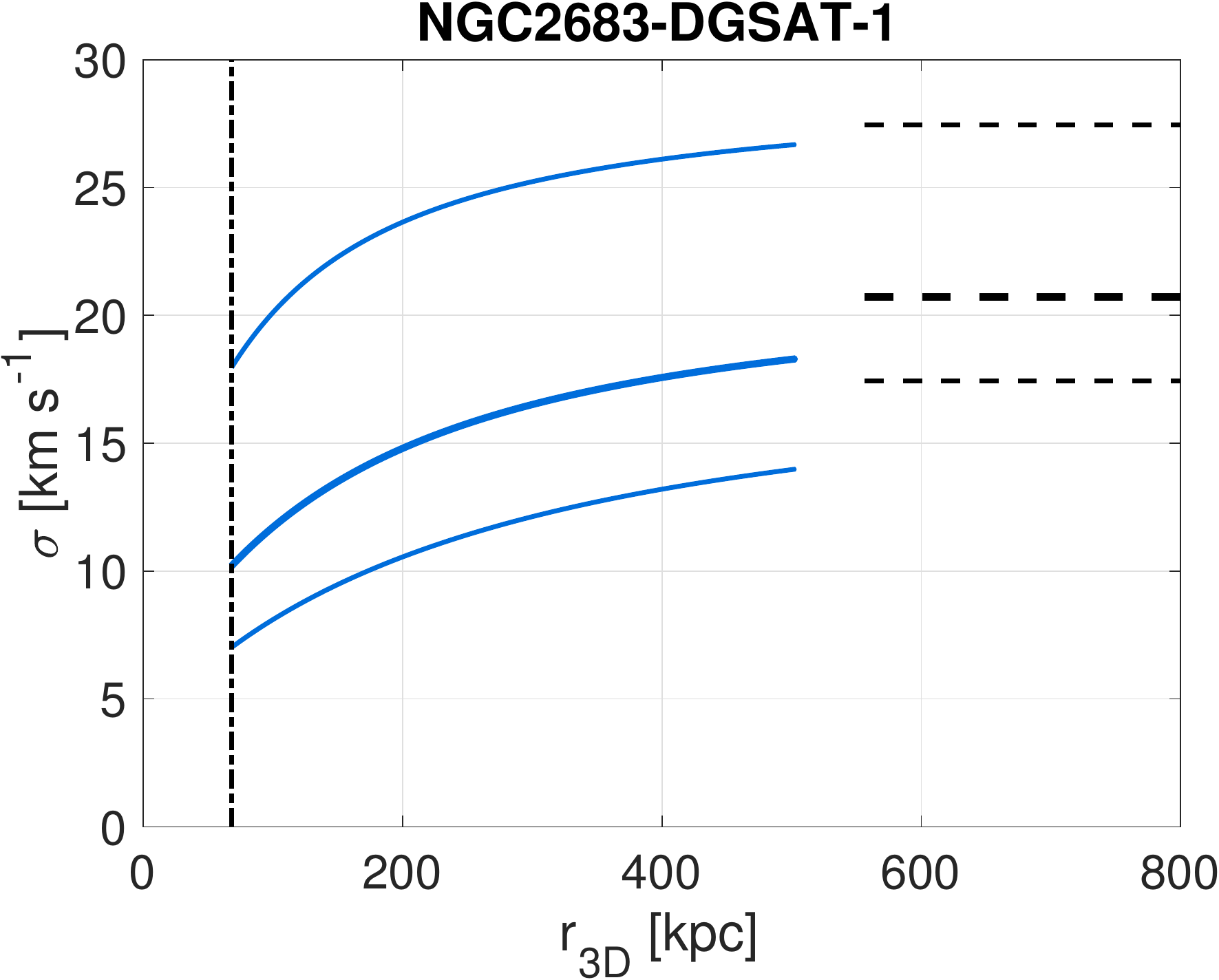}
\includegraphics[width=4.5cm]{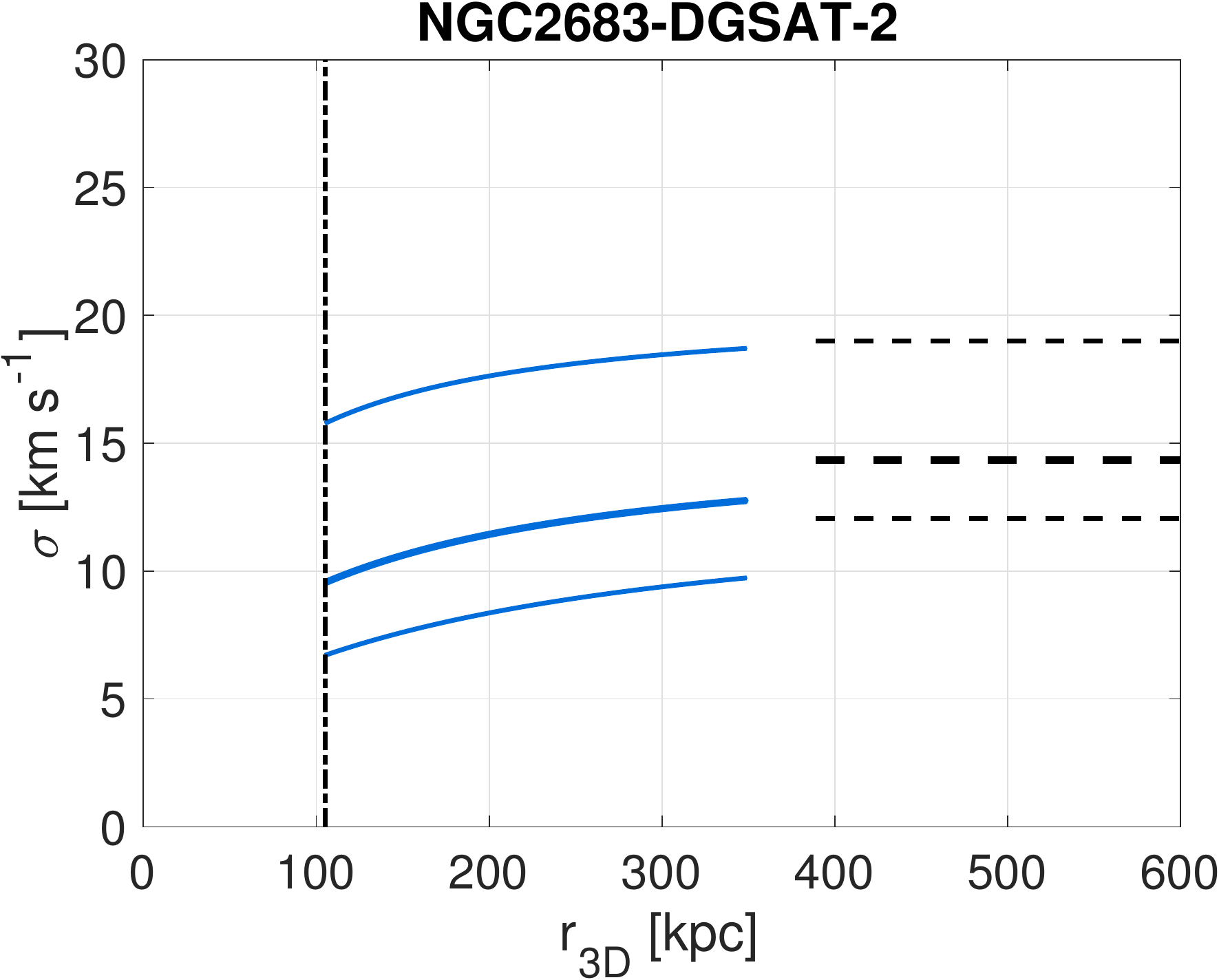}
\includegraphics[width=4.5cm]{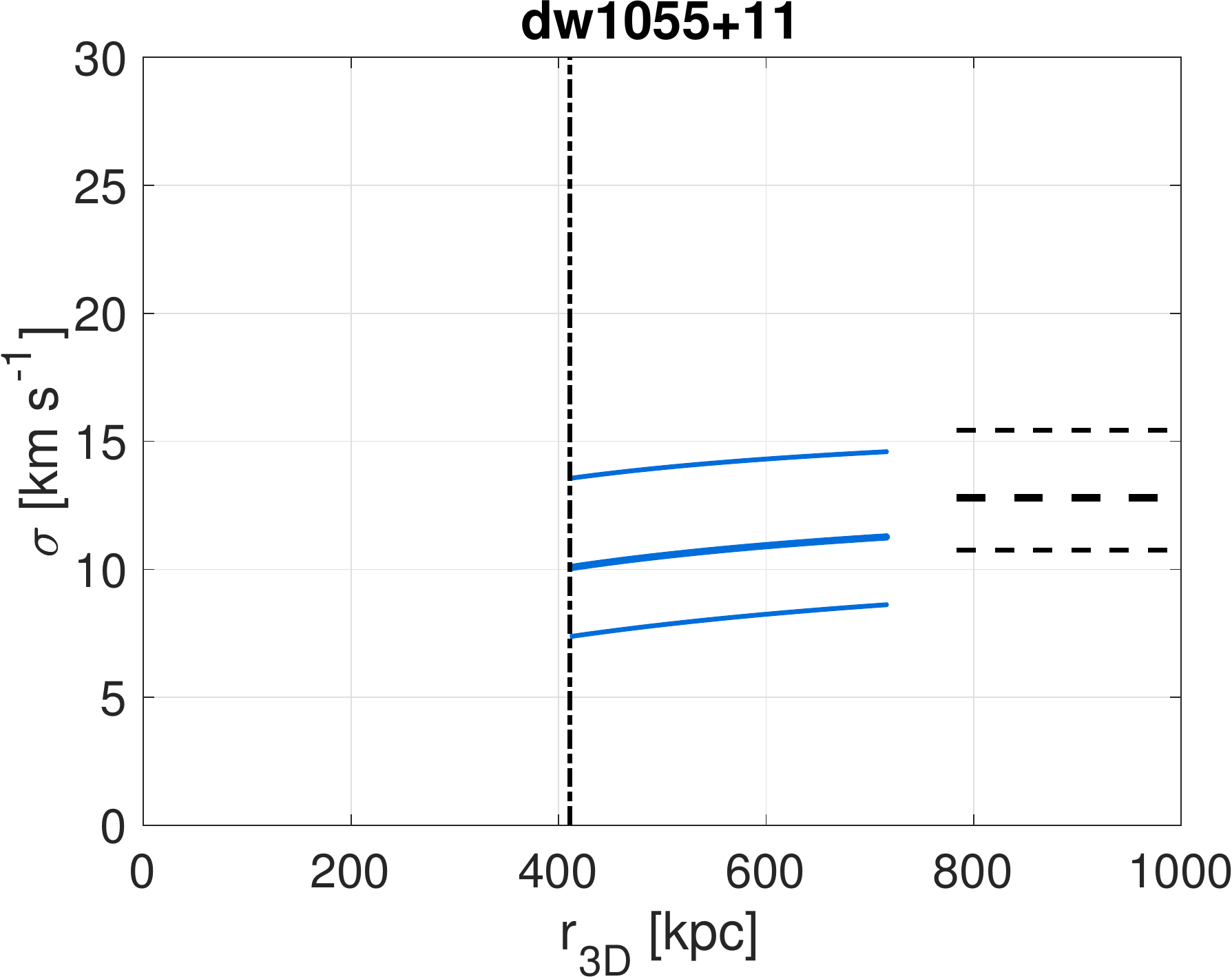}
\includegraphics[width=4.5cm]{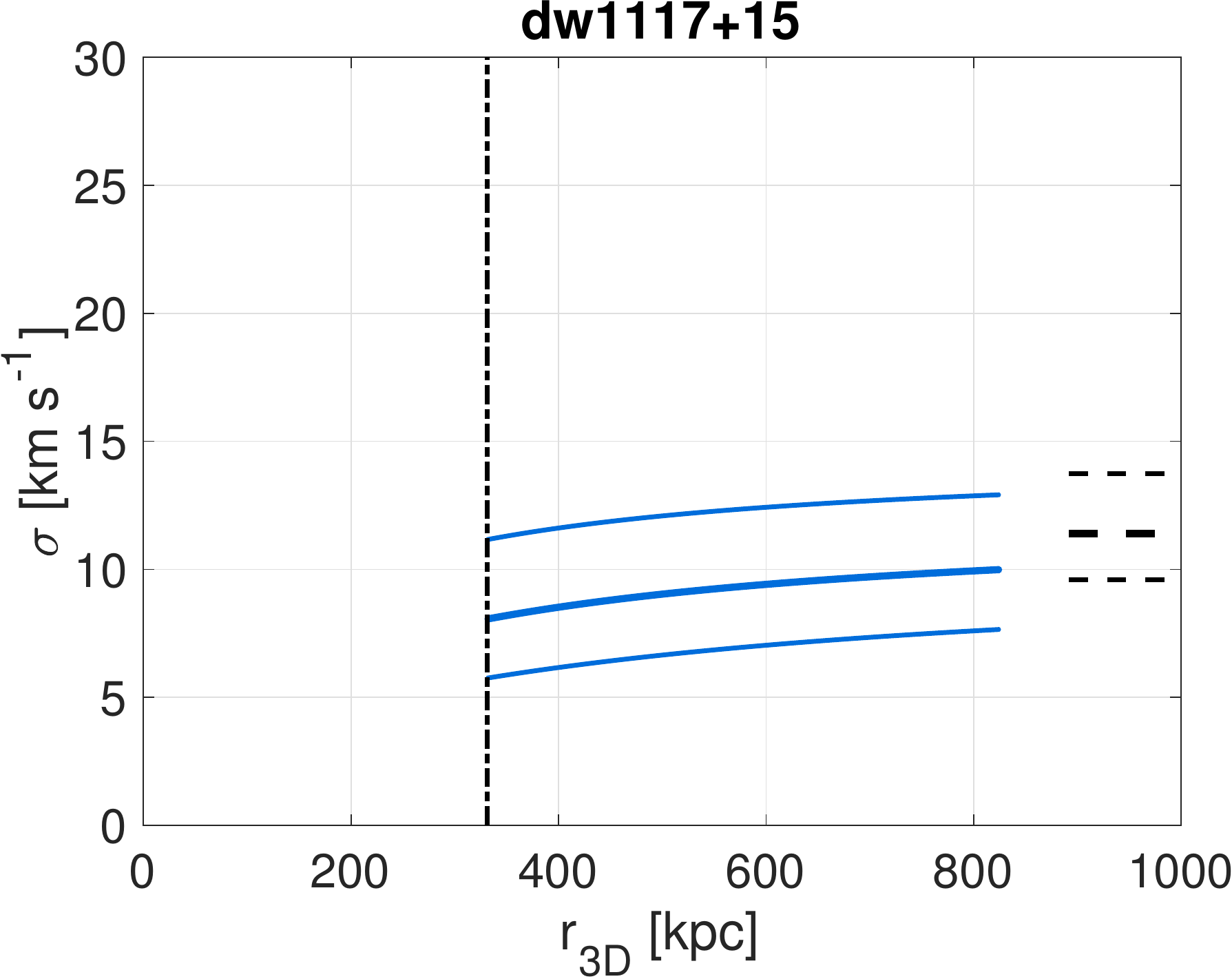}\\
\includegraphics[width=4.5cm]{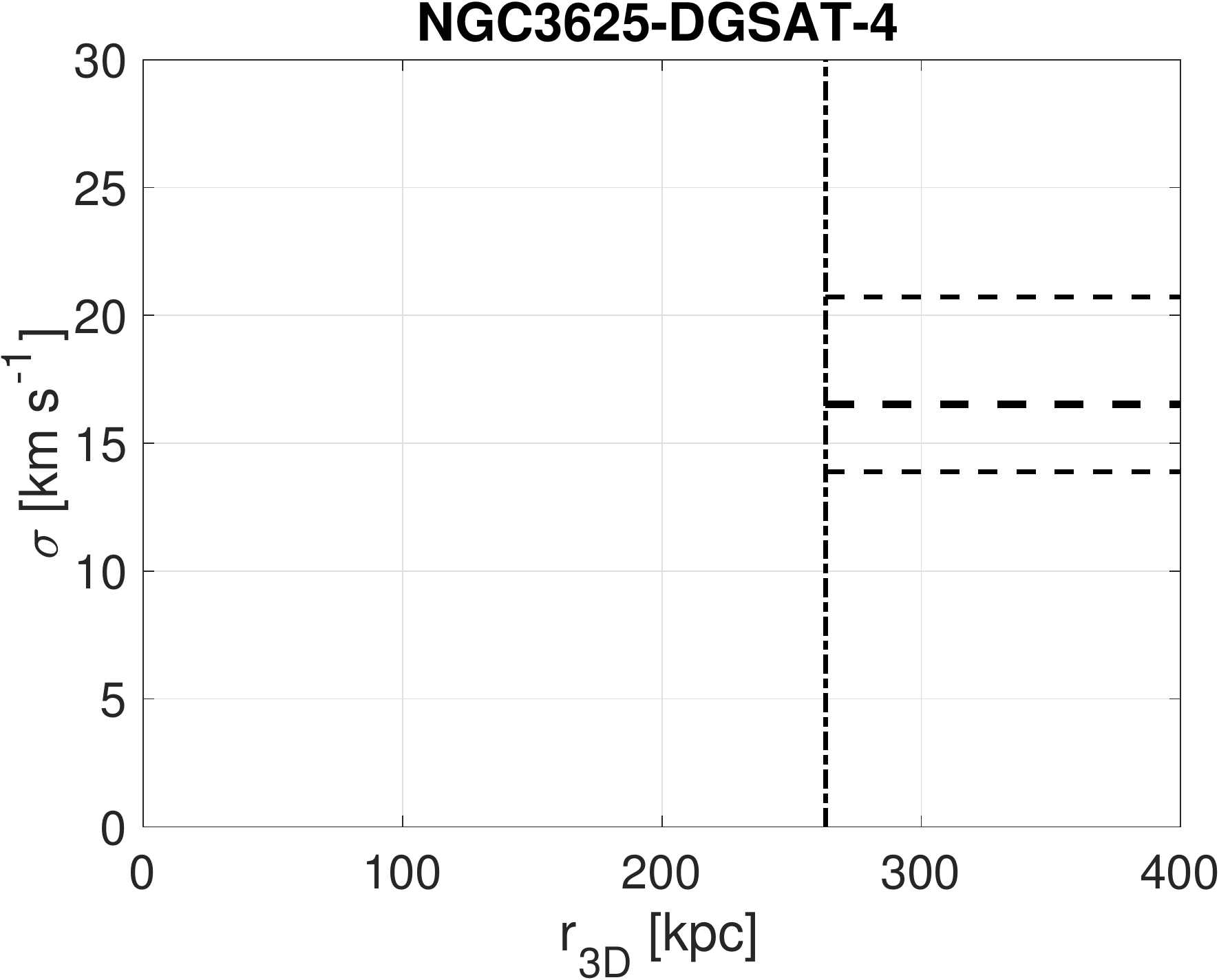}
\includegraphics[width=4.5cm]{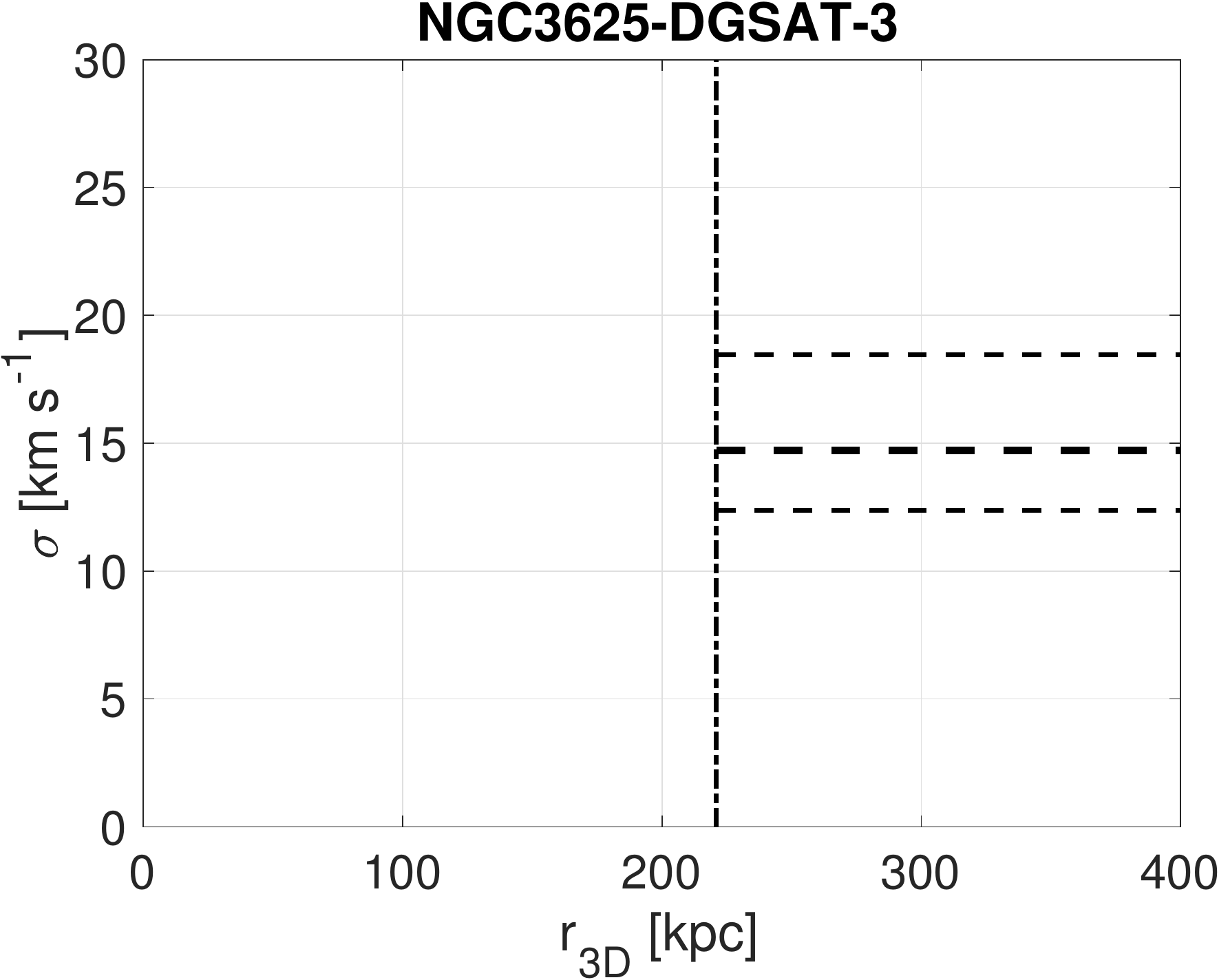}
\includegraphics[width=4.5cm]{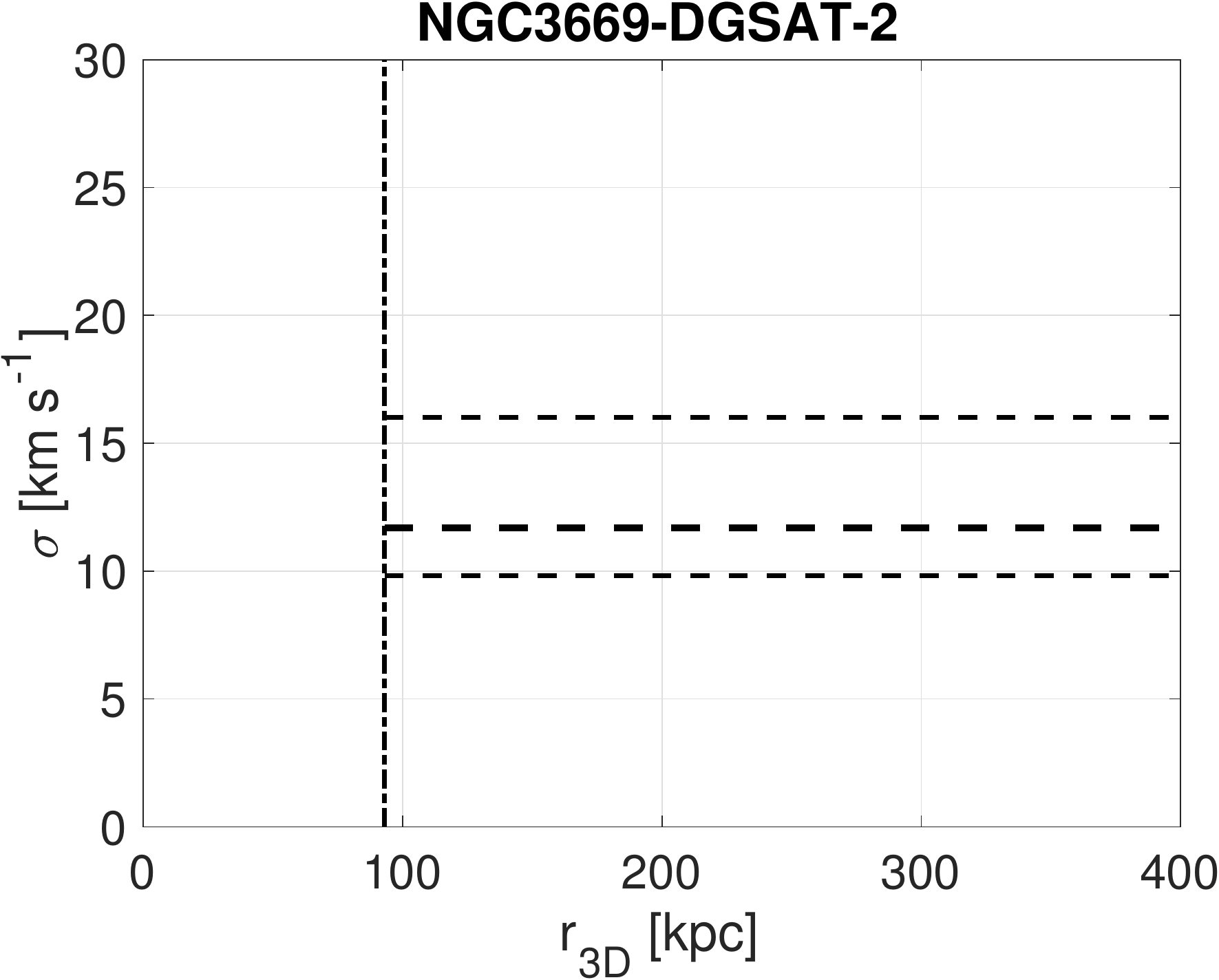}
\includegraphics[width=4.5cm]{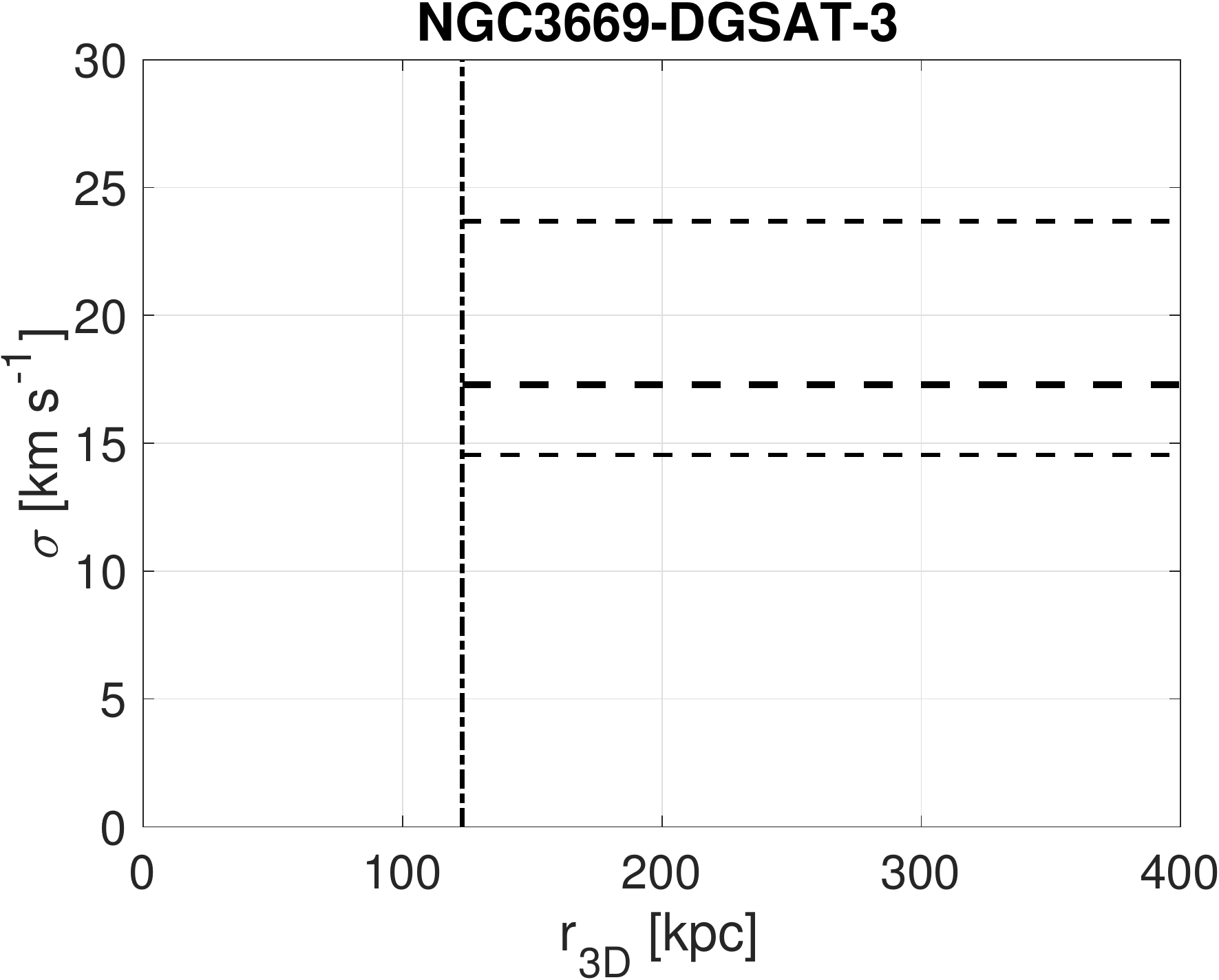}\\
\includegraphics[width=4.5cm]{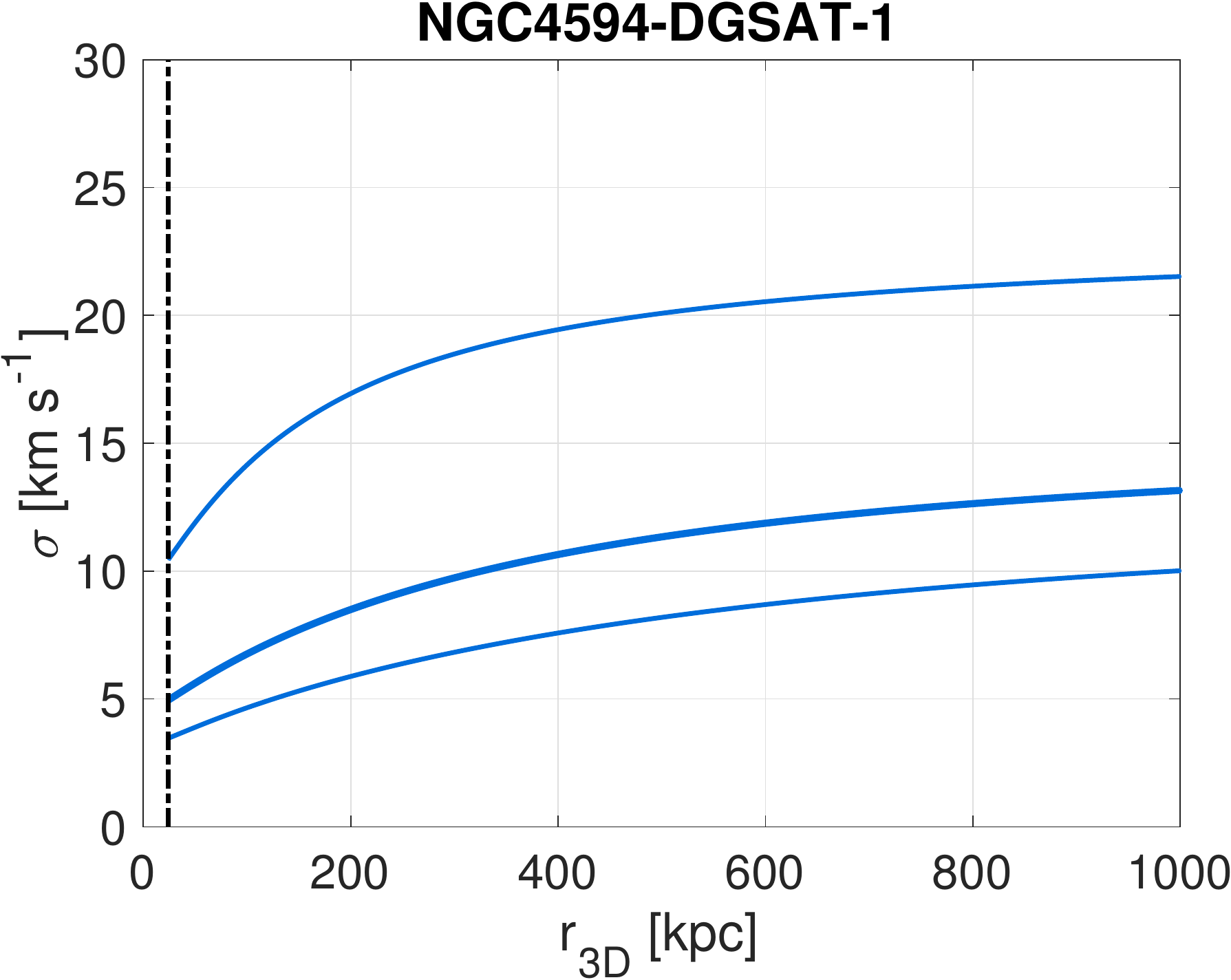}
\includegraphics[width=4.5cm]{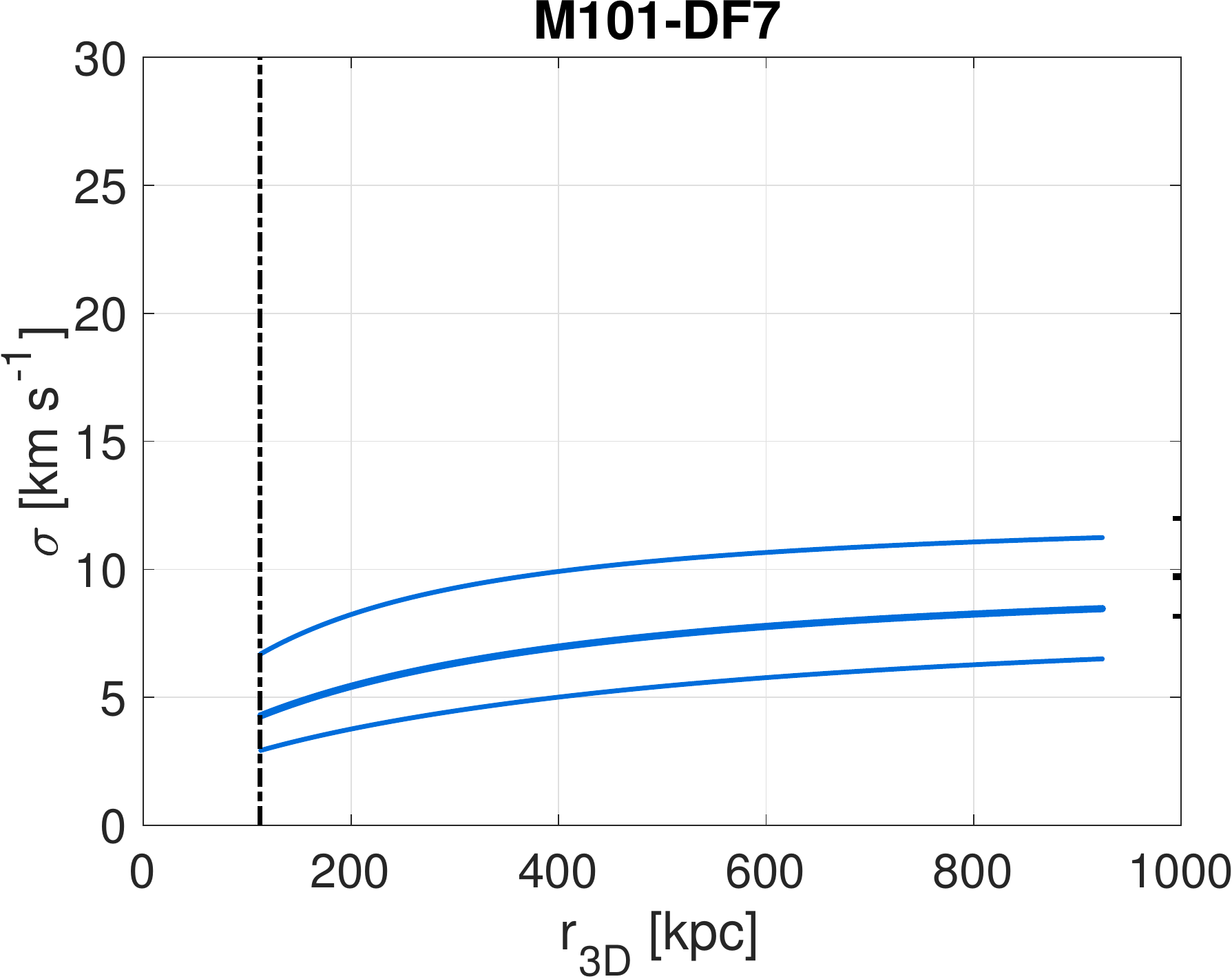}
\includegraphics[width=4.5cm]{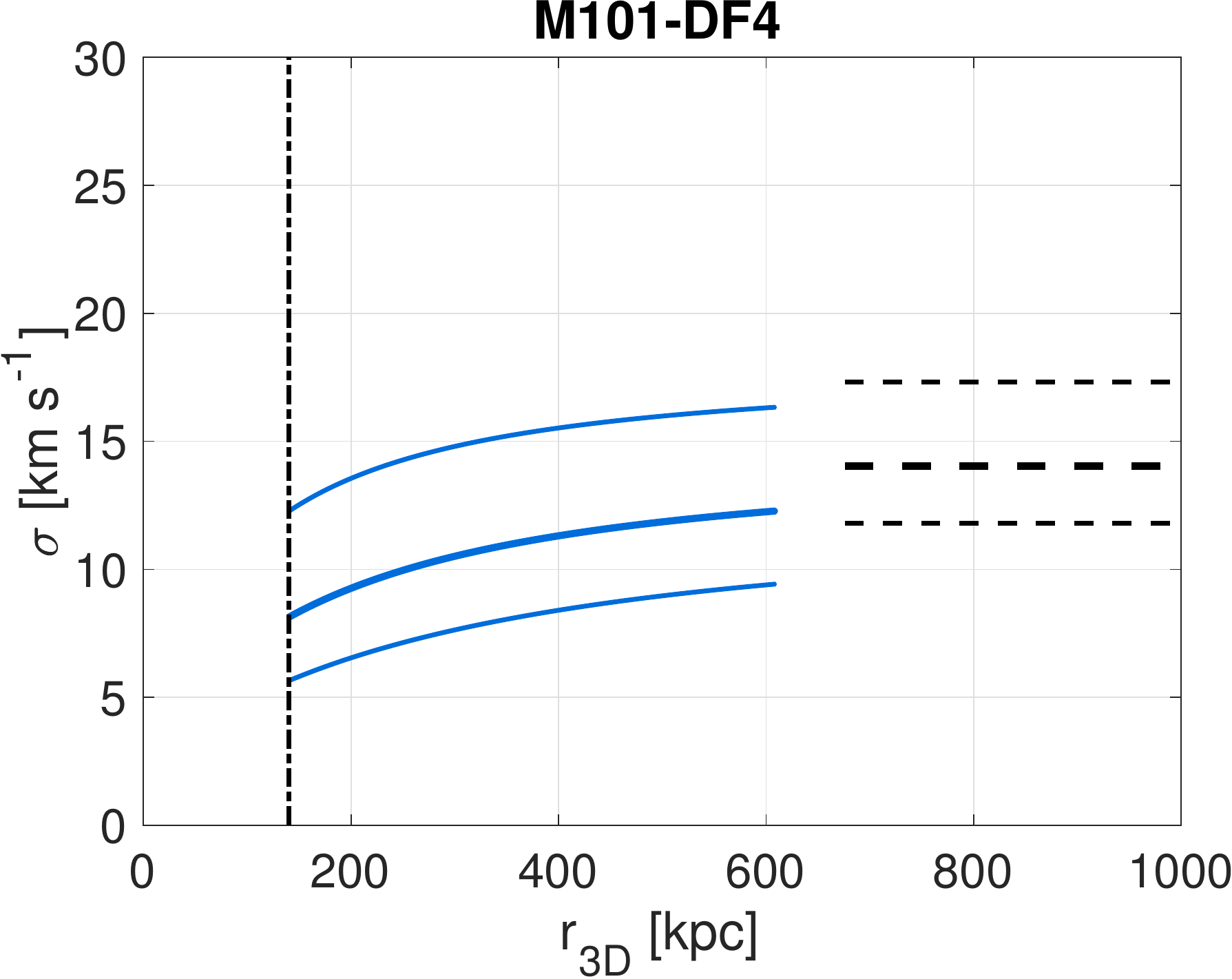}
\includegraphics[width=4.5cm]{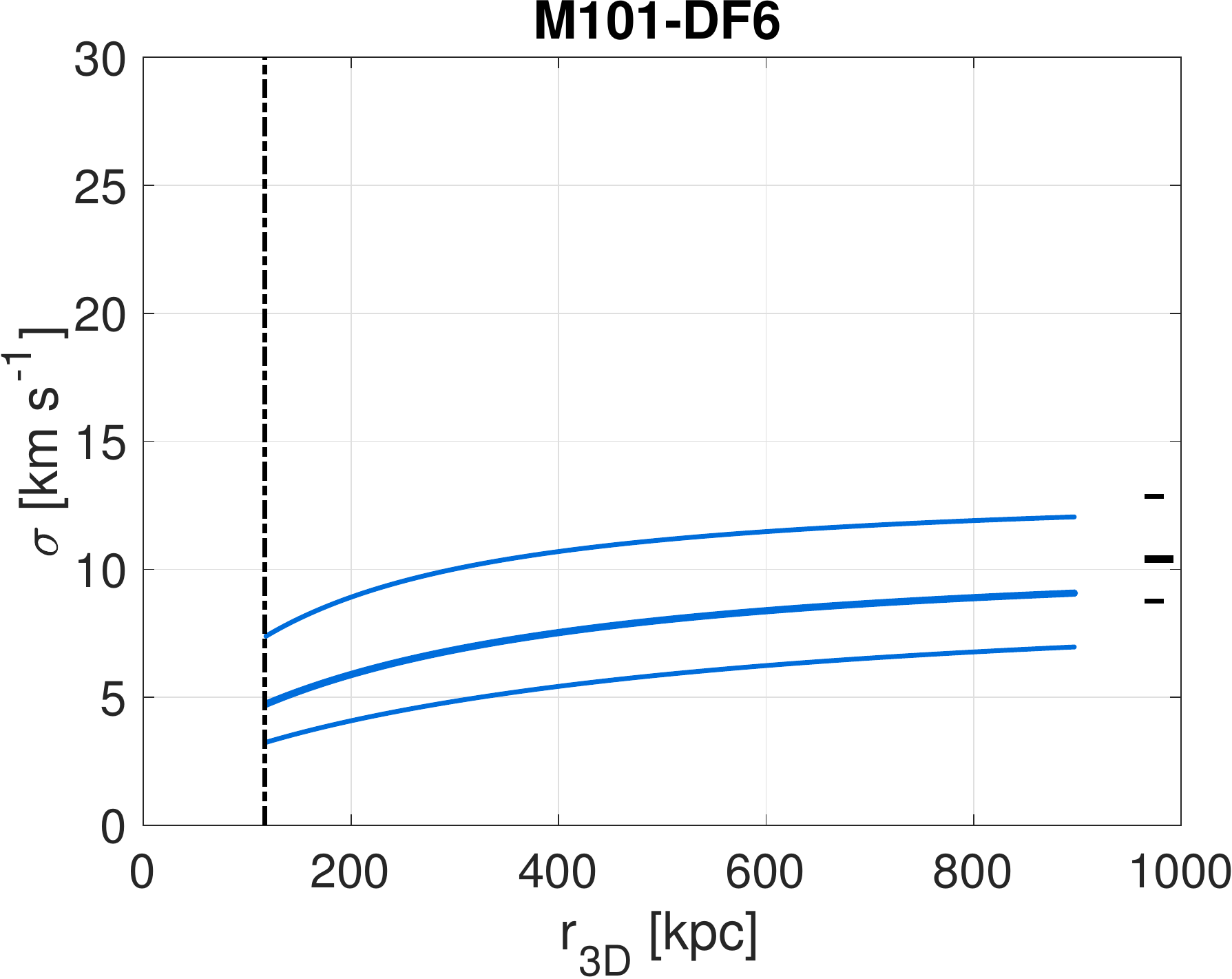}\\
\includegraphics[width=4.5cm]{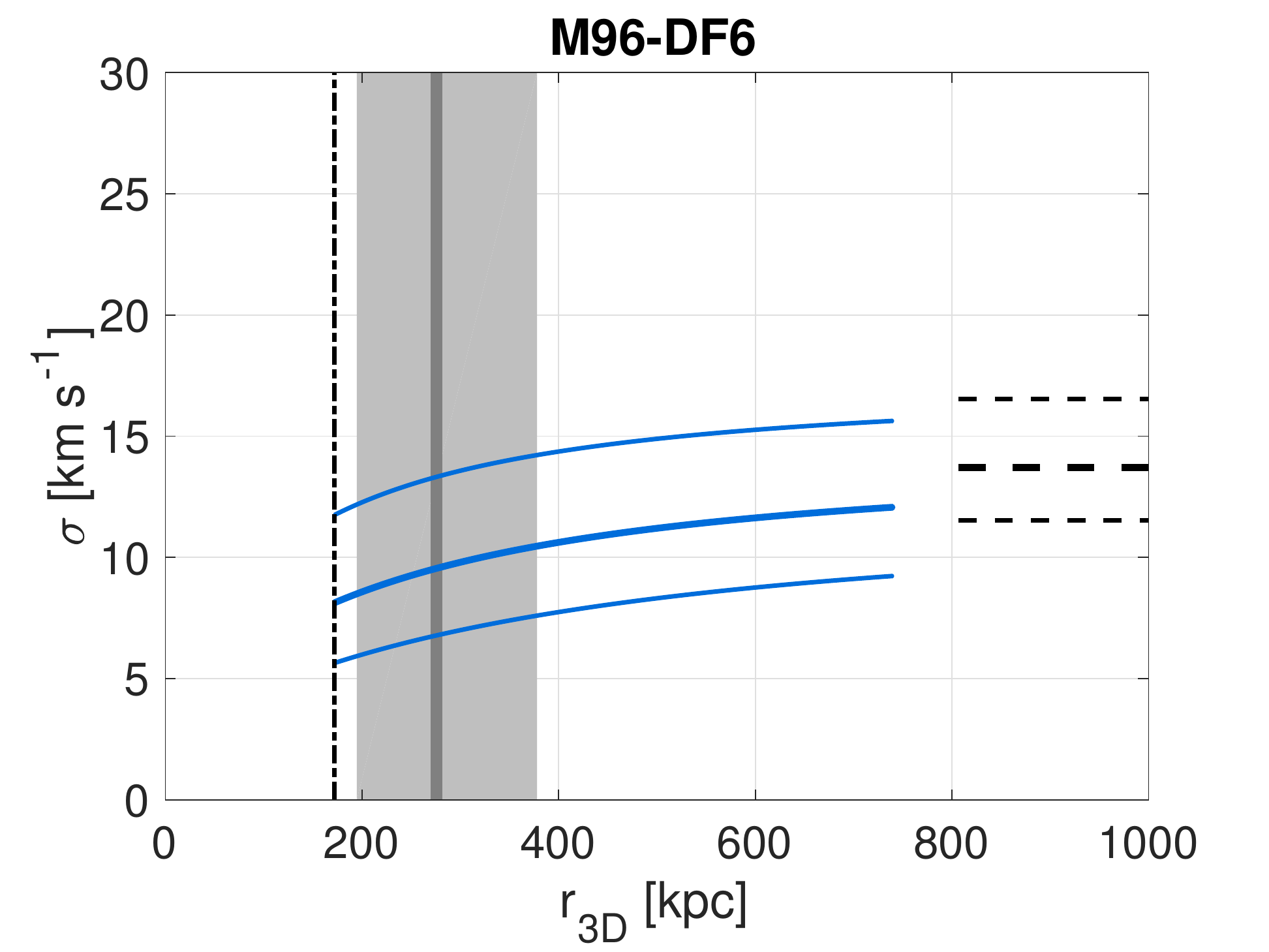}
\includegraphics[width=4.5cm]{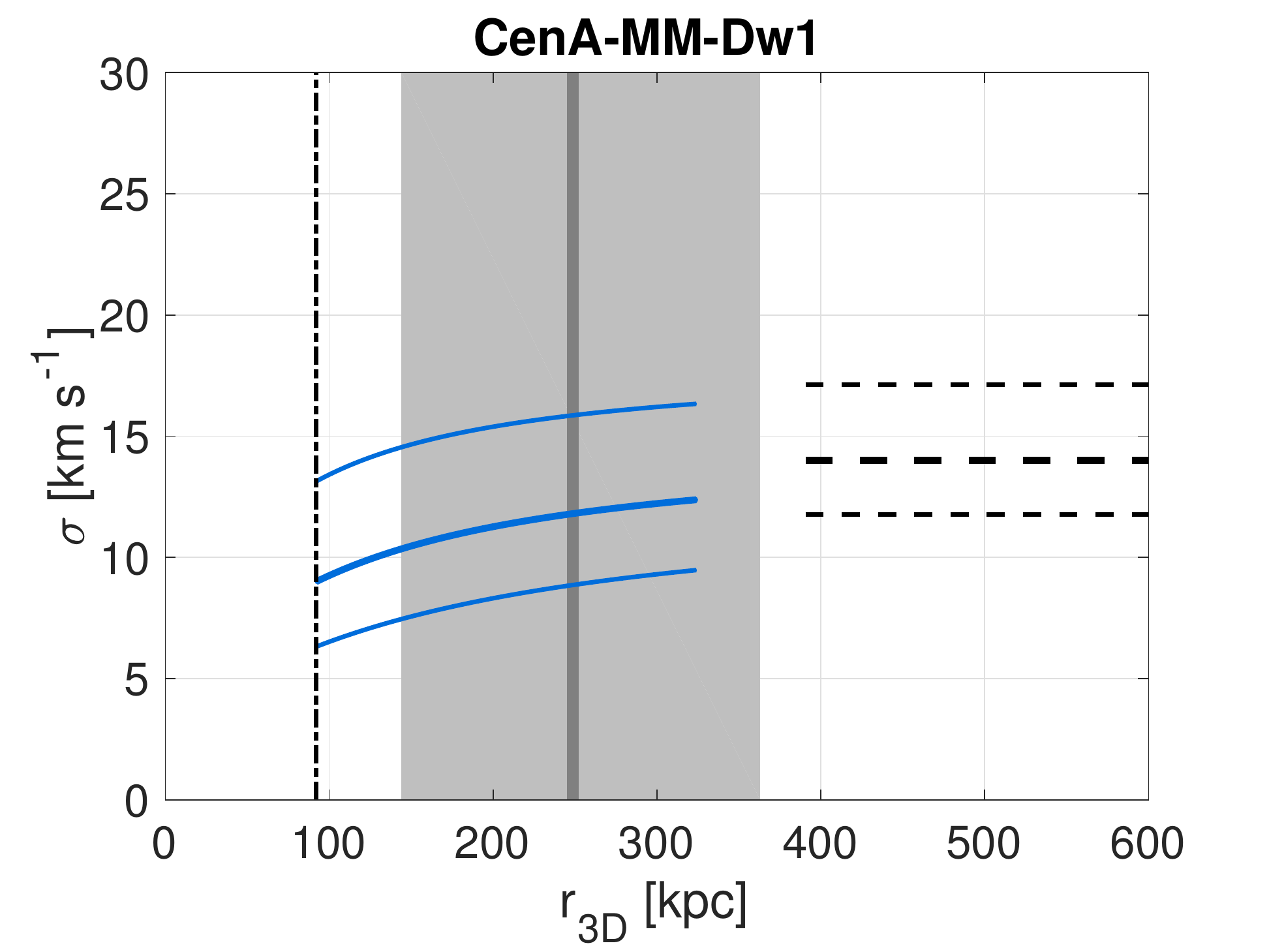}

\caption{The predicted velocity dispersions including the EFE (blue lines) and for the isolated case (dotted lines) as function of the three dimensional separation up to 1~Mpc. The inclusion of the EFE is stopped when $g/g_{ex}>2$. The vertical dotted line corresponds to the on-sky separation (and thereby lower limit of the 3D distance) between the host and the UDG. The upper and lower limits of the curves are derived by varying the $M/L$ and the distance to the host, see text. The last two entries -- M\,96-DF6 and Cen\,A-MM-Dw1 -- have high-precision distance measurements available, we therefore indicate the three dimensional separation (grey line) as well as the associated $1\sigma$ uncertainties (grey area).
}
\label{pred1}
\end{figure*}

  In Figure \ref{pred1} we show the predictions for the 18 UDGs in our sample which have one clear host. The curves start at the projected separation of the UDG to its host. We derive upper and lower limits for the curves as follows: for every UDG, we vary the $V$-band $M/L$ ratio between 1 and 4, as well as take into account the distance uncertainty to the host. We also allow the mass of the host to vary by $\pm20$ percent. The lower limit is therefore given by a $V$ band $M/L=1$, a distance to the host at its lowest limit (making the UDG less luminous), and the $+20$ percent upper limit for the host mass, whilst the upper limit is given by a $V$ band $M/L=4$, the upper limit for the host galaxy system distance (making the UDG more luminous), and the $-20$ percent lower limit for the host mass. We also give a typical estimate in between those two curves, with $M/L=2$ and the literature values for the distance and mass (or rotational velocity) of the host. We have simplified the Figures by using the transition $g/g_{ex}=2$ for the typical case also for the transition for the lower and upper limits. Correctly we would need to take into account that in the lower case, the EFE will have a larger range until the UDG reaches the regime of isolation. We also left a gap between the $\sigma$ estimation including the EFE and the isolated case. This is caused by the fact that we would need to carefully interpolate between the two regimes, which is not the focus of this work.
  
  For 4 UDGs in our catalog, the host dominating the EFE is ambiguous. For NGC\,3625-DGSAT-2 the two possible host galaxies are not even within the same group and have therefore different nominal distances. For the others, to make a precise prediction, one would need to use a numerical Poisson solver as a function of the full 3D configuration of the galaxy group. To give an idea of the range of possible predictions, we provide in Fig.~\ref{predmulti} the predictions as a function of the separation to each potential host galaxy separately.
  
   \begin{table}[ht]
\caption{The predicted velocity dispersions.}% title of Table
     % is used to refer this table in the text
\centering                          % used for centering table
\setlength{\tabcolsep}{2pt} %
\begin{tabular}{l c  l l l l }        % centered columns (4 columns)
\hline\hline                 % inserts double horizontal lines$\alpha_{2000}$ & $\delta_{2000}$ & Observing & Instrument & Exposure & Filter & Airmass & Image quality  \\    % table heading 
Name & $\sigma_{typical}$  & $\sigma_{min}$  & $\sigma_{max}$  \\    % table heading 
 & km s$^{-1}$ & km s$^{-1}$ &km s$^{-1}$  \\ 
\hline      \\[-2mm]                  % inserts single horizontal line
NGC\,7814-DGSAT-2 & {4.6} & 3.0 & 16.2 \\
NGC\,7814-DGSAT-7 & {7.4} & 4.9 & 13.5 \\
NGC\,1052-DF4 & {17.3} & 9.2$^*$ & 23.4 \\
NGC\,1052-DF1 & {6.8} & 3.7$^*$ & 15.8 \\
NGC\,1052-DF2$^1$ & {15.3} & 10.2 & 25.1 \\
NGC\,1084-DF1 & 19.8 & 16.6 & 24.1 \\
NGC\,2683-DGSAT-1 & {10.9} & 7.0 & 27.4 \\
NGC\,2683-DGSAT-2 & {10.1} & 6.7 & 19.0 \\
M\,96-DF6 & {8.7} & 5.6 & 16.5 \\
dw1055+11 & {10.5} & 7.4 & 15.4 \\
dw1117+15 & {8.5} & 5.7 & 13.7 \\
NGC\,3625-DGSAT-2$^2$ & {9.6} & 7.0 & 13.7 \\
 & {6.1} & 4.0 & 11.0 \\
NGC\,3625-DGSAT-4 & 16.5 & 13.9 & 20.7 \\
NGC\,3625-DGSAT-3 & 14.7 & 12.4 & 18.5 \\
NGC\,3669-DGSAT-2 & 11.7 & 9.8 & 16.0 \\
NGC\,3669-DGSAT-3 & 17.3 & 14.5 & 23.7 \\
NGC\,4594-DGSAT-1 & {5.1} & 3.5 & 21.5 \\
Cen\,A-MM-Dw1 & {9.6} & 6.3 & 17.1 \\
M\,101-DF5 & {13.9} & 6.6$^*$ & 20.8 \\
M\,101-DF6 & {5.1} & 3.2 & 12.8 \\
M\,101-DF4 & {8.4} & 5.6 & 17.3 \\
M\,101-DF7 & {4.6} & 2.9 & 12.0 \\

%Name & & & & & & & & \\
\hline
\end{tabular}
\tablefoot{$^*$denotes that potentially the ultra-diffuse galaxy is embedded in multiple strong external fields, in which case we present the smallest velocity dispersion value from all of them.\\$^1$ {The values here are slightly higher than in \cite{2018MNRAS.480..473F} because of very slightly different assumptions for the baryonic mass of the UDG and host, as well as in the definition of the upper limit}. \\ $^2$NGC\,3625-DGSAT-2 could also be associated to NGC\,3619 (second shown entry) instead of NGC\,3625 (first shown entry), which is closer to us, therefore we provide the predictions for both cases.}
\label{list} 
\end{table}

%  \begin{figure*}[ht]
%  \includegraphics[width=6cm]{prediction_NGC1084-DF1-eps-converted-to.pdf}
%  \includegraphics[width=6cm]{prediction_NGC1052-DF2-eps-converted-to.pdf}
%\includegraphics[width=6cm]{prediction_M101-DF7-eps-converted-to.pdf}\\
%\includegraphics[width=6cm]{prediction_M101-DF6-eps-converted-to.pdf}
%\includegraphics[width=6cm]{prediction_M101-DF4-eps-converted-to.pdf}
%\includegraphics[width=6cm]{prediction_dw1117+15-eps-converted-to.pdf}\\
%\includegraphics[width=6cm]{prediction_dw1055+11-eps-converted-to.pdf}
%\includegraphics[width=6cm]{prediction_CenA-MM-Dw1-eps-converted-to.pdf}
%\caption{Continuation of Figure \ref{pred1}.}
%\label{pred2}
%\end{figure*}
  
    \begin{figure*}[ht]
\includegraphics[width=9cm]{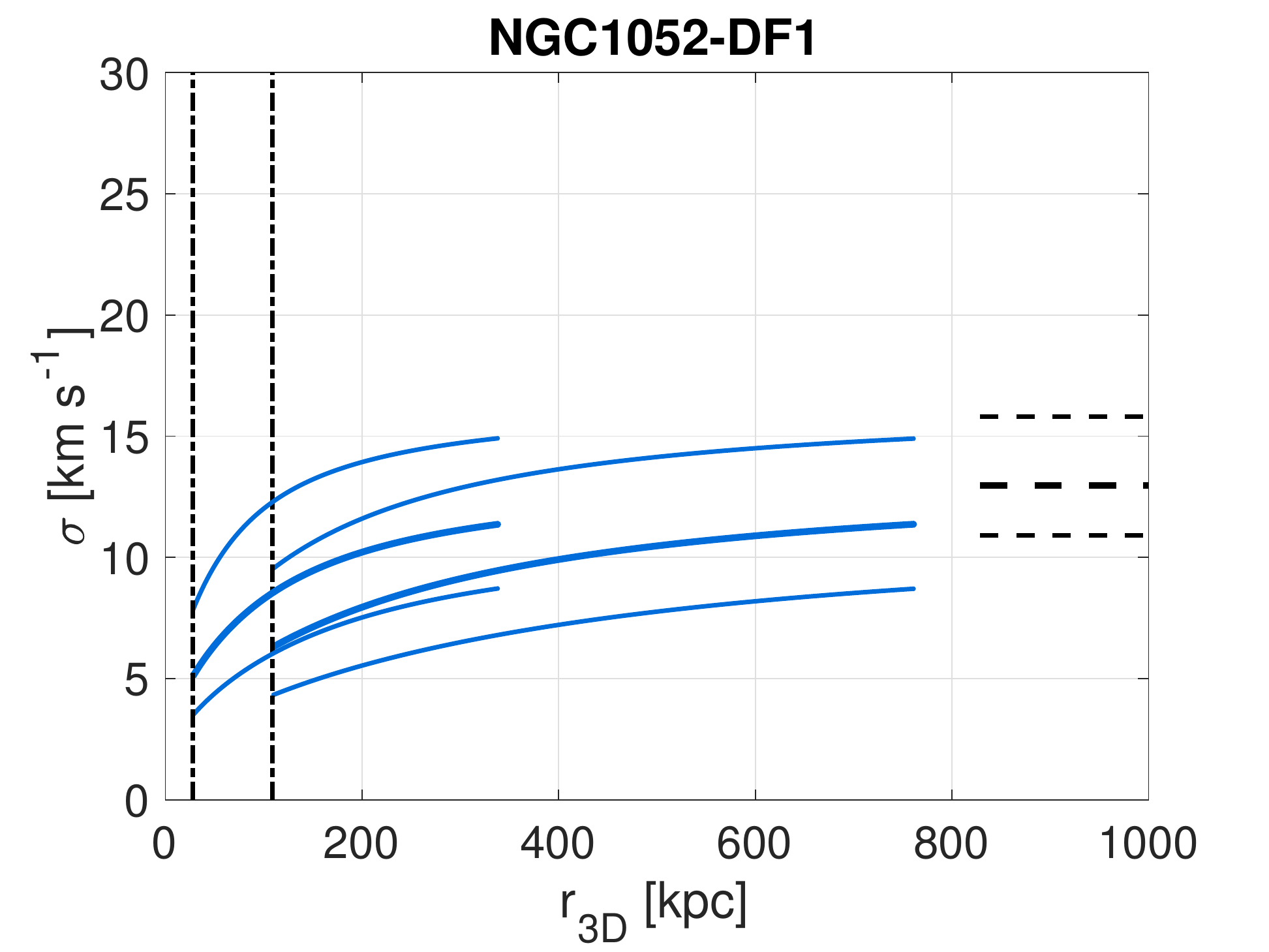}
\includegraphics[width=9cm]{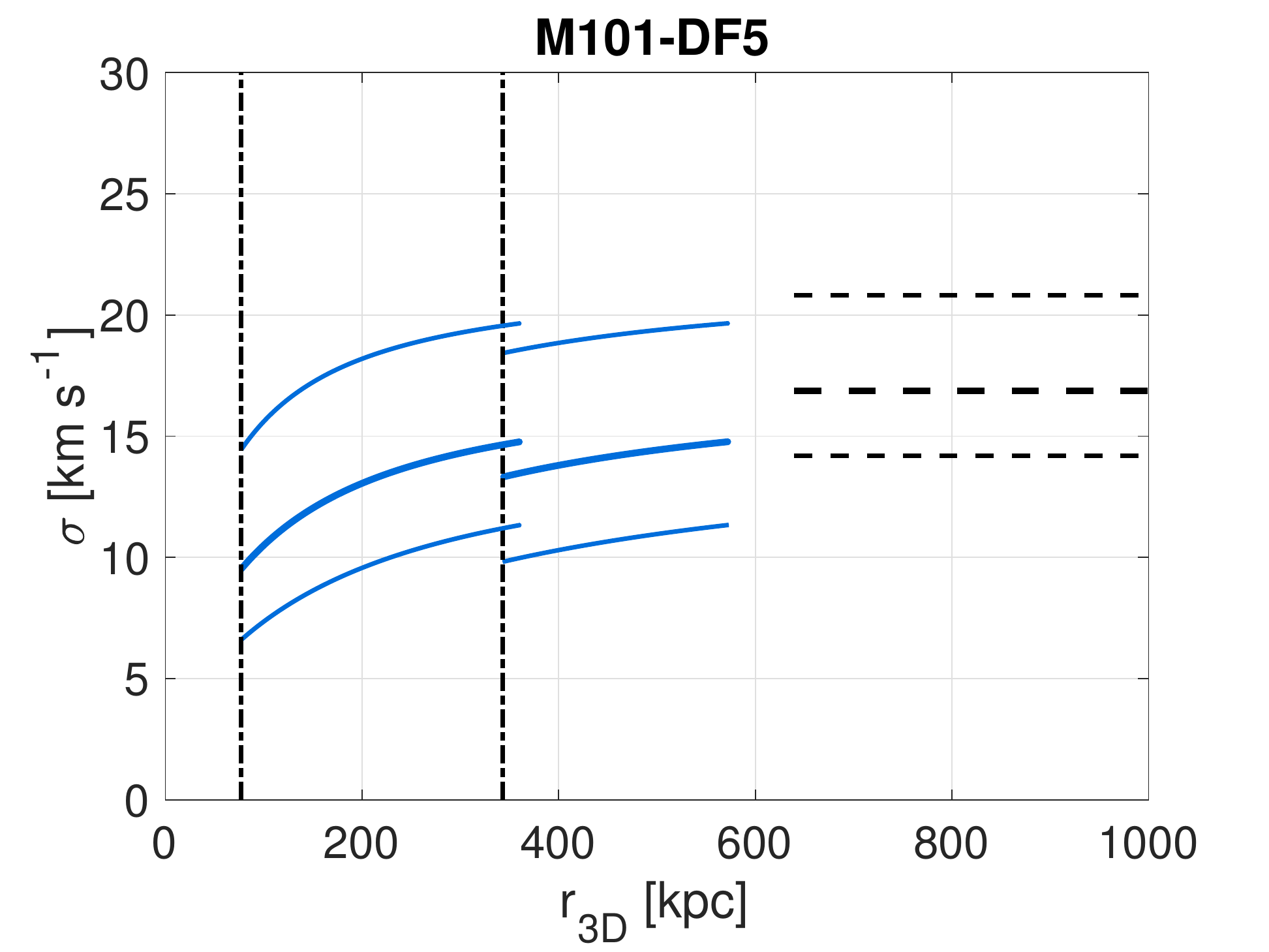}\\
\includegraphics[width=9cm]{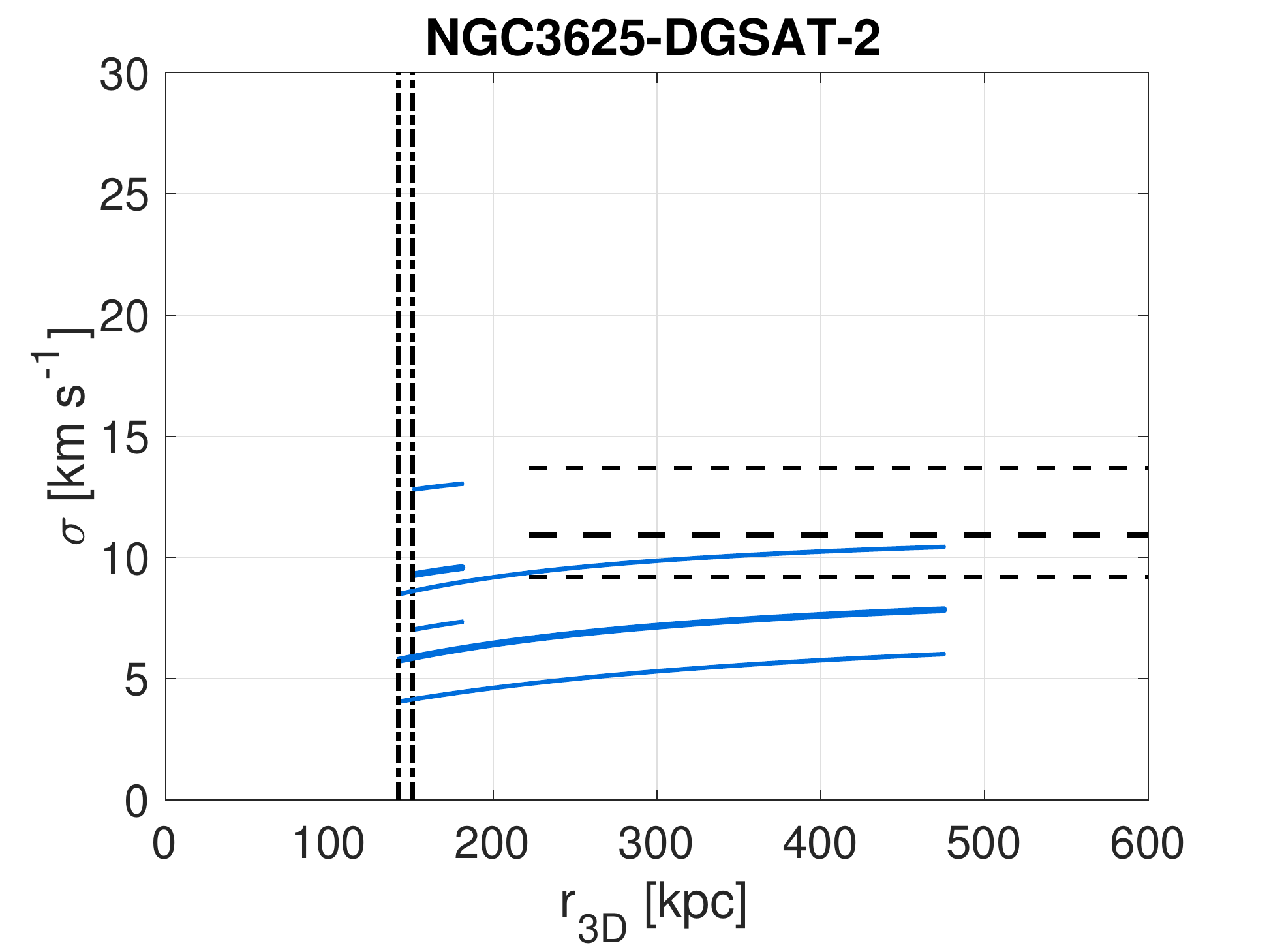}
\includegraphics[width=9cm]{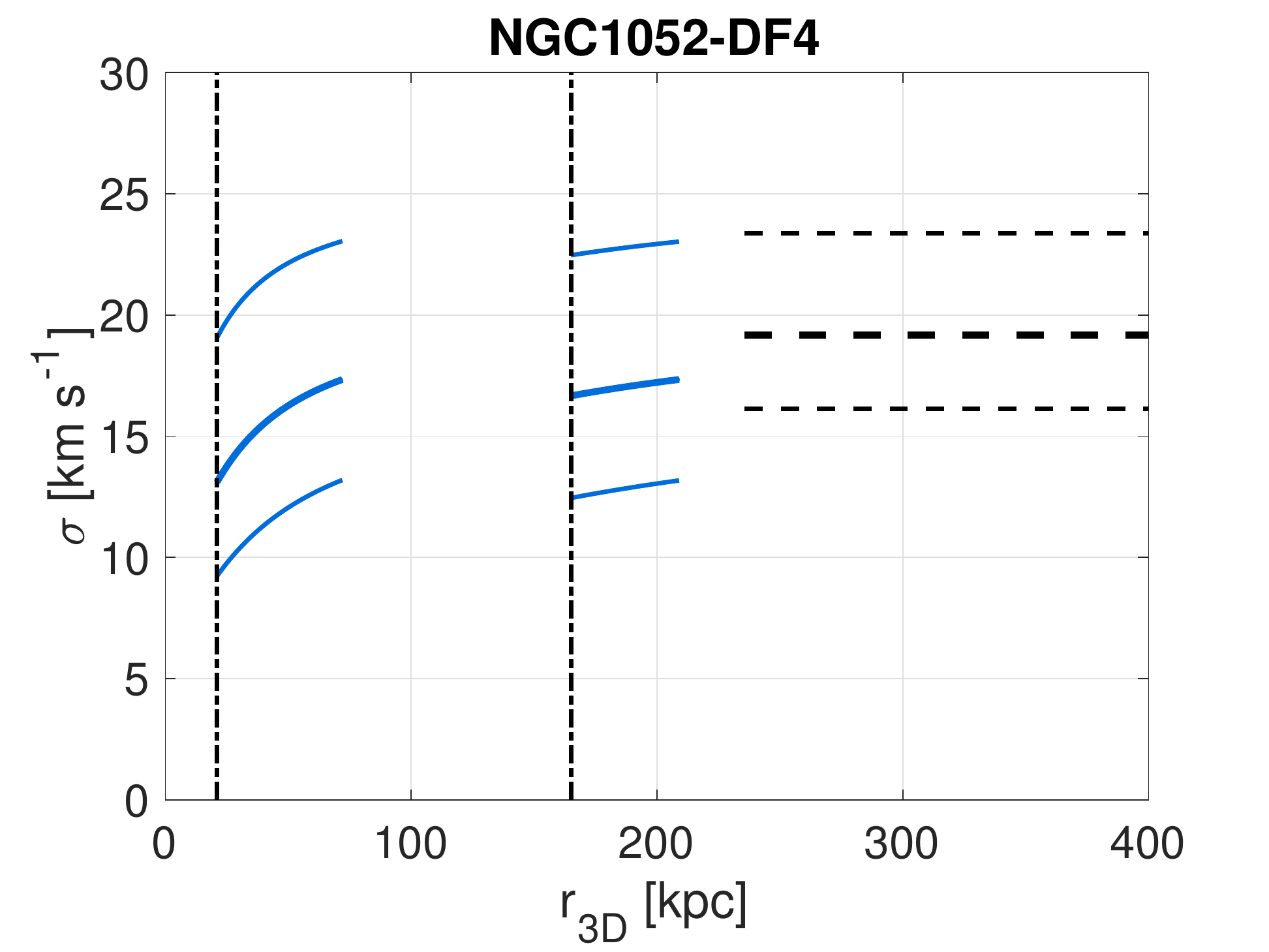}

\caption{Continuation of Figure \ref{pred1}, in the cases where multiple external fields have potentially a strong influence on the ultra-diffuse galaxies. For NGC\,3625-DGSAT-2 the two host galaxies are not within the same group and have therefore different nominal distances.}
\label{predmulti}
\end{figure*}
  
    In Table \ref{list} we present the minimal and maximal allowed values for the velocity dispersion, as well as a typical value for the UDG, if it has a $V$ band $M/L$ ratio of 2 (which is typical for old dwarf galaxies) and has a physical distance of $\sqrt{3/2}\cdot\Delta_{proj}$. This distance simply assumes that the depth along the line-of-sight is the same as the on-sky separation.
  
  For 17 of our 22 UDGs, we note that the EFE was indeed important to consider and will effectively lower the stellar velocity dispersion of the galaxy for most of its possible 3D separations. 
  
  \section{Discussion}
  \label{discussion}
  
      \begin{figure}[ht]
\includegraphics[width=9cm]{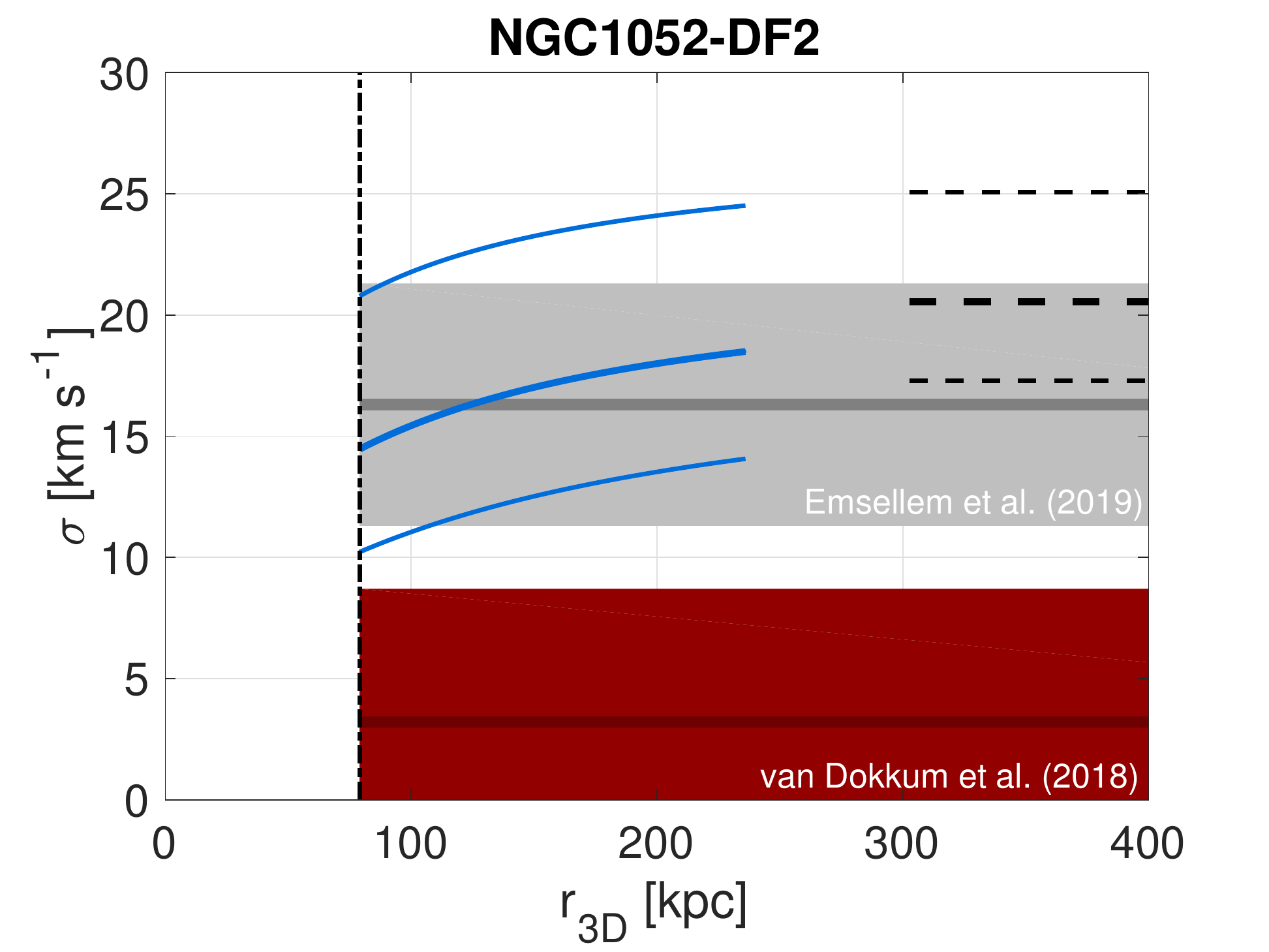}

\caption{The velocity dispersions for the ultra-diffuse galaxy NGC\,1052-DF2. The blue and black lines correspond to the MOND predictions as in Figure \ref{pred1}. The red area denotes the $1\sigma$ range of the velocity dispersion measurement of \citet{2018Natur.555..629V}, the gray are the stellar velocity dispersion measurement of \citet{2018arXiv181207345E}.}
\label{df2}
\end{figure}
  
  The discovery of the UDG NGC\,1052-DF2 with its spectroscopic follow-up observations opened up a novel opportunity to study MOND and the EFE in this class of galaxies. In Figure \ref{df2} we present the MONDian calculation for this galaxy, this time including the $\sigma$ measurements by \citet{2018Natur.555..629V} and \citet{2018arXiv181207345E}. The former gave a velocity dispersion of $\sigma=3.2_{-3.2}^{+5.5}$\,km s$^{-1}$ (shown in red in Figure \ref{df2}), using globular clusters as tracer of the stellar body of the system. It is apparent that such a low value for the velocity dispersion would indeed be an outlier in MOND (to be more precise, a $2\sigma$ outlier, see Figure 1 in \citealt{kroupa2018does}).
\citet{2018arXiv181207345E}, using precise IFU spectroscopy -- obtained with MUSE mounted at the Very Large Telescope -- directly derived  a stellar velocity dispersion of $\sigma=16.3_{-5.0}^{+5.0}$\,km s$^{-1}$ (shown in grey in Figure \ref{df2}). This value is in excellent agreement with the MONDian estimates made here and in previous studies \citep{kroupa2018does,2018MNRAS.480..473F}. Additionally, \citet{2018arXiv181207345E} found two more globular clusters and three Planetary Nebulae. Using them, and updated velocities for five previously known globular clusters, they derive a velocity dispersion of $\sigma=10.5_{-2.2}^{+4.0}$\,km s$^{-1}$, which is at the lower limit of allowed values in MOND (see Table \ref{pred}).

  Assuming that the stellar velocity dispersion $\sigma=16.3$\,km s$^{-1}$ is the true value for this system, and that we live in a MONDian universe, NGC\,1052-DF2 is then most likely at a physical separation of 124\,kpc to its host NGC\,1052, considering the external field. This does make insofar sense, that this is a typical radial separation of a satellite to its host. However, NGC\,1052-DF2's systemic velocity is off by 300\,km s$^{-1}$, which could indicate that this galaxy is not a bound member of the NGC\,1052 group. Only precise distance measurements can tell whether this galaxy indeed belongs to the group, or whether it is a field galaxy in the fore-or background. In the latter case, an estimate with the corresponding distance would be needed to check the validity of MOND.
  
  The MUSE observations are indeed an encouraging prospect for future studies of these UDGs. At these low-surface brightness levels, it is possible to get an estimate of the velocity dispersion under ten hours of observation time, which is expensive but certainly doable. Other today's available facilities which potentially can conduct such studies are KCWI installed at Keck and  Megara at the Gran Canaria Telescope. In the future, the next-generation telescopes like the  ELT will certainly be able to conduct such measurements in short times for a large sample of galaxies.
 
  We refrain here from making any predictions of the velocity dispersion of the UDGs in terms of the standard $\Lambda$CDM framework. In principle, we could collect the dark matter halos in high-resolution simulations for the satellite galaxies, given the observed luminosity and construct a range of possible values for the velocity dispersion. However, the origin of  UDGs is still not understood. If they are failed Milky Way type galaxies, they will possess a large dark matter halo to begin with, and on the other hand, if they origin from dwarf galaxies, they will reside in smaller dark matter halos. This huge range of possibilities makes it unfeasible to make a prediction of the velocity dispersion in the standard context. 
  
  We note that two of the UDGs within our sample -- Cen\,A-MM-Dw1 and M\,96-DF6 --  already have precise distance measurements available \citep{2018arXiv180905103C,2018ApJ...868...96C}.  The three dimensional separation between Cen\,A and Cen\,A-MM-dw1 is $249$\,kpc with a distance uncertainty for Cen\,A-MM-dw1 of $\pm120$\,kpc. This galaxy resides at a separation where the EFE plays a role. In Figure \ref{pred1} we indicate the measured as well as the allowed $1\sigma$ separations derived from the errors. Assuming that Cen-A-MM-Dw1 follows our typical estimates, it should have a velocity dispersion of $\sigma=11.8_{-1.5}^{+2.2}$\,km s$^{-1}$. The upper limit is given by the isolated case. For  M\,96-DF6, the three dimensional separation between M\,96 and M\,96-DF6 is 276\,kpc, with a distance uncertainty for M\,96-DF6 of $\pm300$\,kpc. This gives a typical MONDian velocity dispersion of $\sigma=9.5_{-1.0}^{+1.0}$\,km s$^{-1}$. For these estimates, we have not varied the distance to the host. Once taken into account, this will additionally increase the allowed interval for the velocity dispersion. 
  
 {With the updated velocity dispersion for NGC\,1052-DF2 at hand,}
within the $\Lambda$CDM paradigm it is possible to estimate the total mass (dark matter + baryonic mass) with equation 2 given in \citet{2010MNRAS.406.1220W}. With a stellar velocity dispersion of $\sigma=16.3$\,km s$^{-1}$ we get a mass within $r_{1/2}$ of $5.08\times10^8$\,M$_\odot$. With an absolute magnitude of $-15.3$\,mag in the $V$ band the galaxy has a total baryonic mass of $2.27\times10^8$\,M$_\odot$ (using a $M/L=2$ and a $V$ band magnitude of 4.83 for the Sun), leading to a dynamical to visible mass ratio of 4.5 within the half-light radius. This rough estimate is well in agreement with Local Group dwarfs in the corresponding luminosity regime (see \citealt{2012AJ....144....4M}). Therefore,  the observed velocity dispersion of NGC\,1052-DF2 is not peculiar, and hence comfortable in both paradigms, the $\Lambda$CDM standard model of cosmology and the alternative gravity paradigm (MOND).

If NGC\,1052-DF2 cannot help us to distinguish between MOND and $\Lambda$CDM, are there other galaxies in our sample which do? The most affected galaxy in terms of the EFE  in our sample -- max($\sigma_{iso}/\sigma_{typical}$) -- is NGC\,4594-DGSAT-1. Let us make a gedankenexperiment, where we live in a MONDian universe, but fail to realize that, and thus describe it in terms of Newtonian dynamics. Assume that the calculated typical velocity dispersion correctly describes the UDGs, and thus that NGC\,4594-DGSAT-1 has a typical velocity dispersion of $\sigma=5.8$\,km s$^{-1}$. Now our Newtonian observers have exactly measured this value to infinite precision, and using it with \citet{2010MNRAS.406.1220W} they will derive a mass within $r_{1/2}$ of $5.0\times10^7$\,M$_\odot$. The mass of the baryonic content of the galaxy inferred from the light is $6\times10^7$\,M$_\odot$, giving a dynamical to visible mass ratio within $r_{1/2}$ of only 1.7, and hence a dynamical mass-to-light ratio $M_{dyn}/L$ of 3.4. For a galaxy with  $M_V=-13.9$\,mag this is similar to Milky Way dwarf satellite Leo-I \citep{2008ApJ...675..201M}, sitting at the lower end of the luminosity-$M_{dyn}/L$ relation (see Figure 11 in \citealt{2012AJ....144....4M}). However, if the lowest possible value in MOND was measured, i.e. $\sigma_{min}=3.5$\,km s$^{-1}$, the Newtonian observers would derive a $M_{dyn}/L$ ratio of only 1.2, clearly too small for primordial dwarf galaxies in $\Lambda$CDM. This simple gedankenexperiment shows that we can in principle distinguish between primordial dwarf galaxies in the $\Lambda$CDM framework and the MOND predictions for the cases where the UDG is strongly affected by the EFE and high-precision distance and velocity dispersions measurements are available. {Let us note that such a situation has already happened for some Andromeda dwarfs \citep{2013ApJ...766...22M,2013ApJ...775..139M} and Crater-II \citep{2016ApJ...832L...8M}. As these previous examples show, such an observation} would of course not falsify $\Lambda$CDM, but would pose interesting challenges to it. In terms of falsifying MOND with such objects, as discussed in \cite{2018MNRAS.480..473F}, an issue could always be whether the internal dynamics of a UDG on an eccentric orbit have time to come into equilibrium with the continually changing external field. Hence, once an interesting object challenging MOND is found, elucidating this issue would necessitate more involved simulations. 
  
  On the other hand, to tentatively falsify MOND, it would be well worth to follow-up some of the UDGs for which we have shown that the EFE plays a minor role, or no role at all. In that case, when their velocity dispersion is measured to be lower than the MOND prediction for the isolated case, then there is not much room in MOND for those  velocity dispersions to vary (assuming that the values used for the predictions like the distance and luminosity are correct). This is in contrast to the ultra-faint dwarf galaxies in the Milky Way, which MOND tidal effects would certainly bring out of equilibrium, complicating any comparison between observations and predictions \citep{2010ApJ...722..248M,2018MNRAS.476.3816F,2018arXiv180806634R}.

  \section{Summary and conclusion}
  \label{summary}
  Ultra-diffuse galaxies are puzzling objects for cosmology. While they resemble galaxies like the Milky Way in size, they are as faint and feeble as dwarf galaxies. In MOND -- when they are close to a large galaxy -- they can be affected by its external field effect, making them quasi-Newtonian. This EFE -- unique to MOND -- will lower the velocity dispersion of these systems, when compared to isolation. We have compiled a catalog of UDGs in group environments in the nearby universe to predict their velocity dispersions. We used an analytic expression to include the external field in our calculations and derived  velocity dispersions as function of the true three dimensional separation of the UDGs from their host. When the internal accelerations of the UDGs overrule the external effects, we present the prediction based on simple MOND formula for isolated objects. By varying the Mass-to-Light ratio of the UDG, as well as the distance to its putative host, we give upper and lower limits allowed from MOND. Future observations of these UDGs can therefore test the validity of MOND and test the external field effect. 
  
  Two of our UDGs -- Cen\,A-MM-Dw1 and M\,96-DF6 -- have high-precision distance measurements with the HST. In these cases, we have calculated the three dimensional distances to their host and made a more narrow prediction of the expected velocity dispersion in MOND, assuming that they follow a typical $M/L$ ratio of 2. For Cen\,A-MM-Dw1 we expect a velocity dispersion of $\sigma=11.8_{-1.5}^{+2.2}$\,km s$^{-1}$, and for  M\,96-DF6 a velocity dispersion of $\sigma=9.5_{-1.0}^{+1.0}$\,km s$^{-1}$.
  
  For NGC\,1052-DF2 high-precision IFU data have become  available, taken with VLT+MUSE. This is the first time the stellar velocity dispersion itself of an UDG has been measured. We compared this new measurement to the MOND prediction, already made in \citet{2018MNRAS.480..473F,kroupa2018does}, and find excellent agreement. This is an encouraging prospect, as it shows that testing our predictions is feasible with today's instruments. 
  
  {As a last caveat, we should mention that the predictions made here make rather simple assumptions, such as no strong anisotropy, no rotation on the sky, as well as an analytic MOND estimate with only one interpolating function and with an implicitly small eccentricity for the UDG orbit. As discussed in \cite{2018MNRAS.480..473F}, a UDG on a very eccentric orbit with a continually changing external field could display a lower velocity dispersion than the prediction made here. Hence, once an interesting object challenging MOND is found, elucidating this issue would necessitate more involved simulations.} 
  
\begin{acknowledgements}
    O.M. thanks the Swiss National Science Foundation for financial support. {We thank the referee for their useful and constructive report.} 
\end{acknowledgements}

\bibliographystyle{aa}
\bibliography{bibliographie}

%-------------------------------------------------------------------

\end{document}